\documentclass{article}

\usepackage{arxiv}
\usepackage{url} 
\usepackage{booktabs} 
\usepackage{listings}
\usepackage{xcolor}
\usepackage{amsfonts,amssymb}
\usepackage{graphicx}
\usepackage{textcomp}
\usepackage[htt]{hyphenat}
\usepackage{multirow}
\usepackage{adjustbox}
\usepackage{mathtools}
\usepackage{framed}
\usepackage[most]{tcolorbox}
\usepackage{enumitem}
\usepackage[ruled,vlined]{algorithm2e}
\usepackage[nameinlink]{cleveref}
\usepackage{pifont}
\usepackage[affil-it]{authblk}
\usepackage{subfigure}

\newcommand{\tab}{\hspace*{1em}}

\SetAlFnt{\small}

\begin{document}

\title{Towards Understanding and Demystifying Bitcoin Mixing Services}
\newcommand\CoAuthorMark{\footnotemark[\arabic{footnote}]}

\author[1]{Lei Wu}
\author[1]{Yufeng Hu}
\author[1]{Yajin Zhou\thanks{Corresponding author (yajin\_zhou@zju.edu.cn).}\hspace*{0.4em}}
\author[2]{Haoyu Wang}
\author[3]{Xiapu Luo}
\author[4]{Zhi Wang}
\author[1]{Fan Zhang}
\author[1]{Kui Ren}
\affil[1]{Zhejiang University}
\affil[2]{Beijing University of Posts and Telecommunications}
\affil[3]{The Hong Kong Polytechnic University}
\affil[4]{Florida State University}

\maketitle
\begin{abstract}
One reason for the popularity of Bitcoin is due to its anonymity.
Although several heuristics have been used to break the anonymity,
new approaches are proposed to enhance its anonymity at the same time.
One of them is the mixing service.
Unfortunately, mixing services have been abused to facilitate criminal activities,
e.g., money laundering. As such, there is an urgent need to systematically
understand Bitcoin mixing services.
  
In this paper, we take the first step to understand state-of-the-art Bitcoin mixing services.
Specifically, we propose a generic abstraction model for mixing services
and observe that there are two mixing mechanisms in the wild, i.e. {swapping} and {obfuscating}.
Based on this model, we conduct a transaction-based analysis and successfully
reveal the mixing mechanisms of four representative services.
Besides, we propose a method to identify mixing transactions that leverage the
obfuscating mechanism.
The proposed approach is able to identify over $92$\% of the mixing transactions.
Based on identified transactions, we then estimate the profit of mixing services and
provide a case study of tracing the money flow of stolen Bitcoins. 
\end{abstract}

\section{Introduction}
\label{sec:intro}

Bitcoin~\cite{nakamoto2008bitcoin} has become one of the representative cryptocurrencies.
As of the first quarter in 2020, the total market capitalization of Bitcoin is over $117$ billion US dollars~\cite{coinmarketcap}.
In contrast to traditional payment channels, the decentralization essence of Bitcoin has three characteristics:
1) money can be transferred online directly without the intervention of any third-party banking services; and
2) transactions are verifiable and cannot be reversed; and
3) the \textit{pseudonymity} makes the linkage between Bitcoin addresses and real-world entities hard.
\textit{Anonymity} is regarded as a key factor leading to Bitcoin's popularity~\cite{blau2017price}. 

However, the anonymity is \textit{relationship anonymity}~\cite{pfitzmann2001term},
and can be broken due to the following features of Bitcoin.
First, the complete transaction history is publicly available, namely, the money flow between Bitcoin addresses can be fully revealed.
Second, the mechanism relies on the pseudonymity of addresses used in transactions, which can be broken by aggregating addresses into clusters (or user identities) with heuristics~\cite{harrigan2016clustering} or publicly available data sources~\cite{meiklejohn2013fistful}.
Once address clusters are identified, the complete money flows between clusters (corresponding to different users) can be revealed. 
As a result, the anonymity is no longer preserved.

To improve the anonymity of Bitcoin, several approaches have been proposed.
Some of them aim to hide the transaction information by modifying the Bitcoin protocol
or building additional infrastructures. Such solutions include \textit{Zerocash}~\cite{sasson2014zerocash} and \textit{Monero}~\cite{noether2015monero}.
Others try to set up third-party services to provide enhanced anonymity without modifying the Bitcoin protocol, e.g., \textit{Mixcoin}~\cite{bonneau2014mixcoin} and \textit{Blindcoin}~\cite{blindcoin}.
Corresponding to these approaches, many \textit{altcoins} and \textit{mixing services} emerged.
Although altcoins can achieve stronger anonymity properties~\cite{noether2015monero}, the migration
cost from Bitcoin to altcoins hinders the popularity of altcoins and makes the mixing service
a good alternative choice.

Unfortunately, anonymity is a double-edged sword.
Apart from the benign applications, Bitcoin has been abused as a primary cryptocurrency for criminal activities~\cite{kethineni2018use}, 
including ransomware like \textit{WannaCry}~\cite{bistarelli2018visualizing}, notorious underground markets like \textit{Silk Road}~\cite{christin2013silkroad} and \textit{Ponzi} schemes~\cite{bartoletti2018data}.
Specifically, mixing services are extremely widely used in those activities to facilitate money laundering.
For example, a previous study~\cite{christin2013silkroad} showed that Silk Road extensively uses mixing services.
It has also been reported~\cite{bterloss} that the attacker laundered $7,170$ Bitcoin through \textit{Bitcoin Fog}
(one of the earliest and most famous mixing services), after attacking \textit{Bter.com} (a former Chinese cryptocurrency exchange).
In addition, on May 8, 2019, cryptocurrency exchange giant \textit{Binance} reported that it has suffered from a large scale security breach, resulting in the loss of around $7,074$ BTC (about 40 million dollars at that time)~\cite{binancereport}. 
Further investigation indicated that a large portion of stolen Bitcoins were sent to \textit{Chipmixer}~\cite{clainreport}, a popular mixing service provider.

The extensive use of mixing services makes it difficult to trace suspicious money flow,
as they deliberately obfuscate the relationship between senders and recipients.
Although there is an urgent need to demystify the mixing services, only a few previous works have been published.
For example, the authors performed a simple graph analysis based on data collected from experiments of
selected mixing services~\cite{moser2013mixing}, while
others focused on security issues of mixing services themselves~\cite{de2017analysis}.
In short, there lacks a comprehensive understanding of Bitcoin mixing services.

\noindent
\textbf{Our approach.}\tab
In this paper, we take the first step to systematically study Bitcoin mixing services. 
Our goal is to understand mixing services in a comprehensive way.

To facilitate our analysis, we first propose a three-phase model to depict the workflow of mixing services.
Our study suggests that most mixing services share the same procedure but differ in the \textit{mixing mechanisms}.
Based on this abstraction model, we categorize state-of-the-art mixing mechanisms into
two types, namely, \textit{swapping} and \textit{obfuscating}.

We then conduct an empirical study to analyze mixing services based on real Bitcoin transactions.
To this end, four representative mixing services are selected and analyzed.
Then we collect sample transactions for each service to analyze the mixing mechanisms.
Finally, we propose a heuristic-based algorithm to identify the mixing transactions of the mixing
services with the obfuscating mechanism.

\noindent
\textbf{Results.}\tab
We apply the approach to analyze four representative Bitcoin mixing services, i.e., \textit{Chipmixer}~\cite{chipmixer}, \textit{Wasabi Wallet}~\cite{wasabiwallet}, \textit{ShapeShift}~\cite{shapeshift}, and \textit{Bitmix.biz}~\cite{bitmixbiz}.

For Chipmixer and Bitmix.biz, we interact with these services by sending
Bitcoins to them to collect sample transactions (inputs to the service and outputs from the service).
We conduct $10$ experiments with $4$ inputs to Chipmixer and $6$ inputs to Bitmix.biz.
In total, we collected $8$ and $14$ outputs from them, respectively.
For ShapeShift and Wasabi Wallet, we are able to reconstruct mixing records using provided public APIs. 
Accordingly, we collected $4,850$ mixing transactions from Wasabi Wallet, and $27,411$ cryptocurrency convert
records from ShapeShift.

Based on the collected sample transactions, we conduct a transaction-based analysis to first
determine the mixing mechanism they used, and then reveal their workflow.
Then, we perform an advanced analysis for services using the obfuscating mechanism to identify mixing transactions.
The evaluation result demonstrates that the proposed algorithm is able to identify most (over \textbf{92\%}) of
the mixing transactions. We further estimate the profit of these services, and use a real attack to demonstrate the
capability of our approach to trace the stolen Bitcoins that have been mixed.

\noindent
\textbf{Contributions.}\tab
In summary, this paper makes the following main contributions.
\begin{itemize}
    \item We proposed an abstraction model and approach to systematically demystify state-of-the-art Bitcoin mixing services.
    \item We applied the proposed approach to four representative Bitcoin mixing services, and successfully revealed
    the mixing mechanisms and workflows of these services.
    \item We proposed an advanced analysis to effectively reveal mixing services that employ the obfuscating mechanism by identifying \textit{most} (over $92$\%) mixing transactions. The evaluation results demonstrated the effectiveness of our approach.
\end{itemize}

To engage the community, the dataset of this study is released
at the following link~\footnote{\url{https://github.com/blocksecteam/bitcoinmixing}}.

\begin{figure}[t]
	\centering
	\includegraphics[width=.45\textwidth]{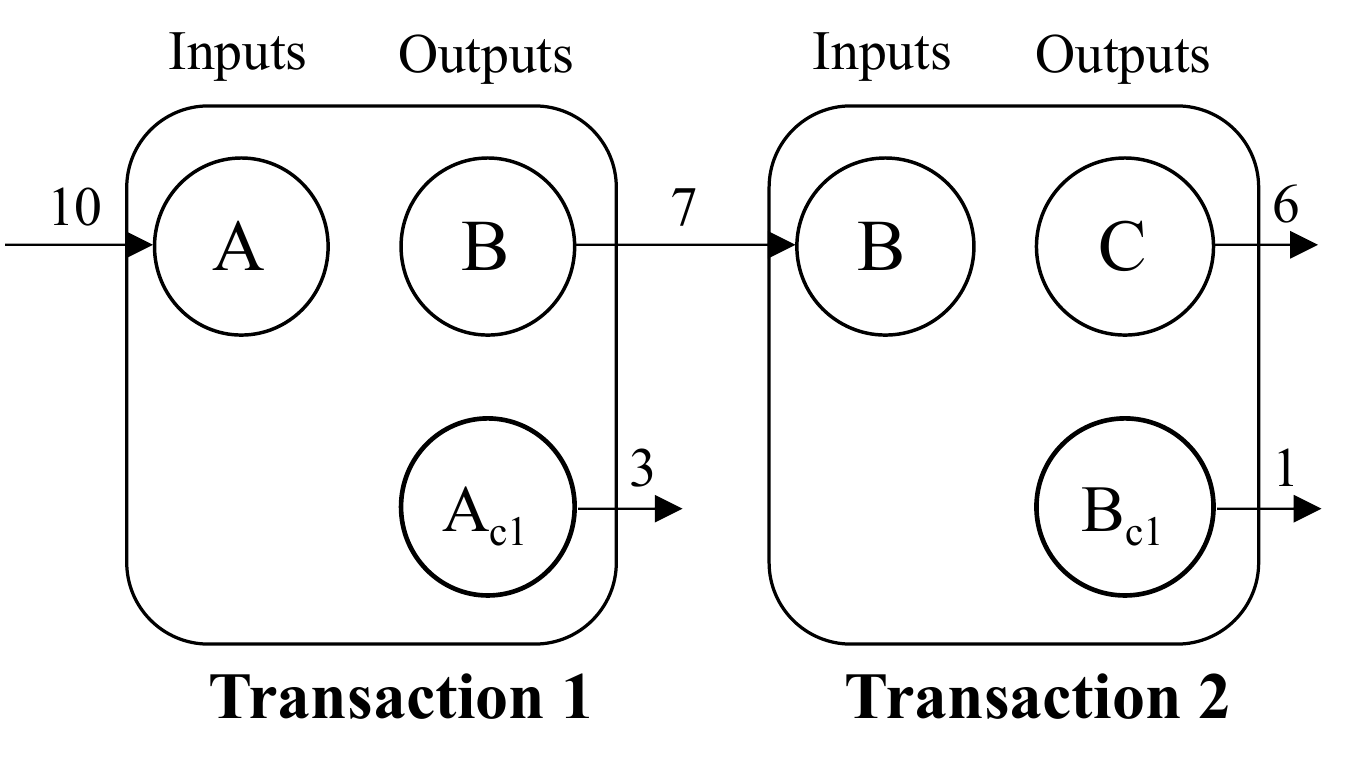}
	\caption{Example Bitcoin transactions.} 
	\label{fig:utxo}
\end{figure}

\begin{figure*}[t]
	\centering
	\includegraphics[width=.9\textwidth]{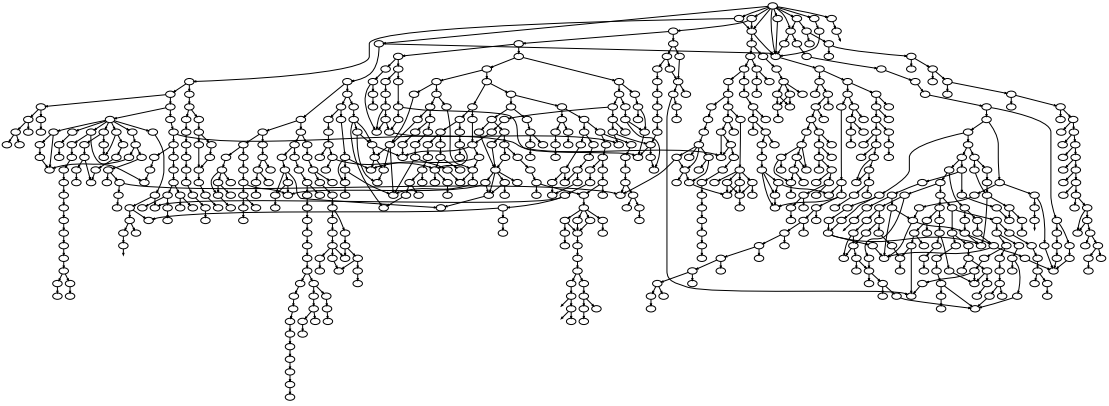}
	\caption{
		The simplified transaction graph for the Binance May Hack case.
		Nodes represent transactions. An edge means the money flow between transactions.} 
	\label{fig:cm_case_raw}
\end{figure*}

\section{Background}
\label{sec:bg}

\subsection{Bitcoin}
\label{subsec:bgbitcoin}

Bitcoin is a decentralized cryptocurrency proposed by an identity with pseudonym Satoshi Nakamoto~\cite{nakamoto2008bitcoin}.
The idea behind Bitcoin is a publicly available and verifiable distributed ledger.
To protect the integrity of this public ledger, Bitcoin employs the Proof-of-Work (PoW) consensus algorithm. 

\noindent
\textbf{Transaction.}\tab 
A transaction is a basic unit describing money flow from input addresses to
output addresses.
Every input is a reference to an {unspent transaction output} (UTXO)~\cite{UTXO}, which is an output in a previous transaction that has not been referenced in other transactions.

Figure~\ref{fig:utxo} gives an example of Bitcoin transactions and UTXOs.
Alice has $10$ BTC in address $A$ (as a UTXO) and wants to send $7$ BTC to address $B$ belonging to Bob.
To this end, Alice initiates a transaction (Transaction 1) referring this UTXO as the input, and specifies two outputs: address $B$ with $7$ BTC and a change address $A_{c1}$ with $3$ BTC.
All outputs in Transaction 1 become UTXOs before they are referenced by other transactions.
Likewise, to send $6$ BTC to address $C$ belonging to Charlie, Bob initiates Transaction 2 referring to the UTXO generated in Transaction 1 as the input, and specifies outputs accordingly.

A transaction is to fully spend UTXOs specified in inputs,
and distribute the remaining one to its output addresses with specified values.
Note that in order to make this transaction verified and confirmed by the Bitcoin network, additional information that verifies the ownership of each UTXO and the integrity of the whole transaction is included in the transaction.
Besides, to broadcast a transaction in the P2P network, users pay {network fees} to the miners who spend the computational resources to verify transactions. 

\noindent
\textbf{Addresses.}\tab 
There are three types of \textit{addresses} in Bitcoin.
Addresses calculated directly from private keys (using hash functions) are called Pay-to-Public-Key-Hash (P2PKH) addresses. They begin with the number prefix \texttt{1}.
In 2012, a new type of address called Pay-to-Script-Hash (P2PSH) was
introduced to simplify the redeem script in transaction output 
for the multiple signature (MultiSig) protocol, and these addresses begin with the
number prefix  \texttt{3}.
In 2017, another new type was introduced in Bitcoin as the segregated witness (SegWit) to separate witness data (to verify the ownership of UTXOs) in transaction inputs. These addresses begin with the prefix \texttt{bc1q}.

\subsection{Mixing Service}
Originating from the Bitcoin community~\cite{bitlaundry,bitcoinlaundry,bitcoinfog},
the underlying idea for \textit{mixing} is to obfuscate the relationship between inputs and outputs, thereby preserving the \textit{relationship anonymity}.

\noindent
\textbf{Centralized Mixing Service.}\tab
A mixing service is called a centralized mixing service if it relies on a central mixing server to perform the mixing.
Many mixing services, such as Bitcoin Fog~\cite{bitcoinfog}, are centralized mixing services.
However, the centralized mixing service
has the \textit{trust} issue.
First, there is no guarantee that the services providers
will send the mixed coins to addresses specified by users.
Second, they can record the \textit{original} relationship
between user inputs and outputs.
Thus, if the services themselves are compromised, the anonymity will be broken.
Mostly due to these reasons, many centralized mixing services
disappeared in recent years, including BestMixer~\cite{bestmixertakedown}, Helix~\cite{helixclosedown} and BitMixer~\cite{bitmixershutdown}.

\noindent
\textbf{Decentralized Mixing Service.}\tab
The decentralized mixing service does not rely on a centralized server to perform the mixing.
{CoinJoin}~\cite{maxwell2013coinjoin} is a generic {decentralized} mixing protocol proposed by Bitcoin Core developers~\footnote{CoinJoin can be implemented in centralized mixing services as well~\cite{khalilov2018survey}.}.
The basic idea is to exploit the structure of transactions to combine different inputs and outputs in a single transaction, thus the recovery of the relationship between
outputs and inputs is becoming harder.
A number of works have been proposed on the basis of CoinJoin, including {CoinShuffle}~\cite{ruffingPedro2014esorics} and {SecureCoin}~\cite{ibrahim2017ijns}.

\noindent
\textbf{Cross-Blockchain Mixing Service.}\tab
There is also a special type of mixing services provided by cryptocurrency exchanges or converters (e.g., ShapeShift~\cite{shapeshift}~\footnote{In this paper, we also study ShapeShift to understand its mixing mechanism. However, we only focus on mixing activities within the Bitcoin network.}, Changelly~\cite{changelly} and Flyp.me~\cite{flypme}). These services allow users to exchange Bitcoin with other cryptocurrencies, e.g., Zcash and Ether.
Obviously, tracking the money flow across different ledgers is not trivial.

\section{Abstraction Model for Mixing Mechanisms}
\label{sec:abs}

As introduced in Section~\ref{sec:bg}, the basic idea of mixing is to hide
relationships between senders and recipients (inputs and outputs), to provide \textit{relationship anonymity}~\cite{moser2013mixing}.
In this section, we propose an abstraction model by separating the mixing process
into three steps, and illustrate the mixing mechanisms.

\subsection{A Motivating Example}

We use a real attack called the \textit{Binance May Hack}~\cite{binancereport}, as the motivating example to demonstrate the difficulty to trace the money flow associated with mixing services.
According to the official announcement of Binance~\cite{binancereport}, the attacker stole ${7,074}$ BTC and withdrew them in one transaction~\footnote{The transaction hash is \texttt{e8b406091959700dbffcff30a60b190133721e5c39e89bb5fe\\23c5a554ab05ea}, and we will use \texttt{e8b406} to denote this transaction in the following.}.
The stolen Bitcoins were then distributed using Chipmixer to perform the money laundering. 
Figure~\ref{fig:cm_case_raw} gives a simplified transaction graph of this attack. 
Specifically, the root node (i.e., the topmost node) represents the withdrawal transaction initiated by the attacker to transfer the stolen Bitcoins, and the subsequent graph shows the tainted money flow through multiple transactions.

Obviously, the graph in Figure~\ref{fig:cm_case_raw} is too
complicated to distinguish mixing transactions from others.
To solve this issue, we propose a general abstraction model in this section
and perform analysis on mixing services in Section~\ref{sec:method}.

\subsection{Mixing in Three Phases}
\label{subsec:3phrases}
The process of a mixing service can be modeled as a three-phase procedure, i.e., \textit{taking inputs}, \textit{performing mixing} and \textit{sending outputs}.
Formally,
a \textit{Mixing Service} (denoted as $\mathcal{S}$) can be defined as a triplet: $(\mathcal{I}, \mathcal{O}, \mathcal{M})$, where $\mathcal{I}$ and $\mathcal{O}$ represent \textit{inputs} and \textit{outputs}, respectively, while $\mathcal{M}$ means the \textit{mixing mechanism}.

Specifically, the mixing service $\mathcal{S}$ first takes Bitcoins to be mixed as the inputs ($\mathcal{I}$).
This is achieved mostly by requiring users to send $\mathcal{I}$ to a service-provided deposit address.
After taking $\mathcal{I}$, $\mathcal{S}$ is responsible for performing mixing with its mixing mechanism ($\mathcal{M}$), which consumes the collected user inputs, and prepares the desired outputs ($\mathcal{O}$) for each user.
Finally, $\mathcal{S}$ will send $\mathcal{O}$ to the users.
Typically, users specify some output addresses to $\mathcal{S}$ to indicate where the mixing output should be sent.

The procedure to handle $\mathcal{I}$ and $\mathcal{O}$ is similar in different mixing services.
In the following, we will focus on different types of $\mathcal{M}$.

\subsection{Mixing Mechanisms}
\label{subsec:mixingmechanisms}
The relationship anonymity is mainly achieved by mixing mechanisms. According to different implementations, they can be further categorized into two types, i.e., \textit{swapping} and \textit{obfuscating}.

To make it more clear, we first give the following definitions:
\begin{itemize}
    \item $\mathcal{M}_S$, the \textit{swapping mechanism};
    \item $\mathcal{M}_O$, the \textit{obfuscating mechanism};
    \item $T_{\mathcal{M}}$, a \textit{mixing transaction}~\footnote{In this paper, we call the transactions created by mixing services as \textit{mixing transactions}.};
    \item $A_N$, an \textit{anonymity set}~\footnote{Groups of outputs with the same value are called \textit{anonymity sets}.} with capacity N ($N \geq 2$), i.e., it has $N$ outputs in a transaction with the same value. 
\end{itemize}

\begin{figure}[t]
\centering
\includegraphics[width=.7\textwidth]{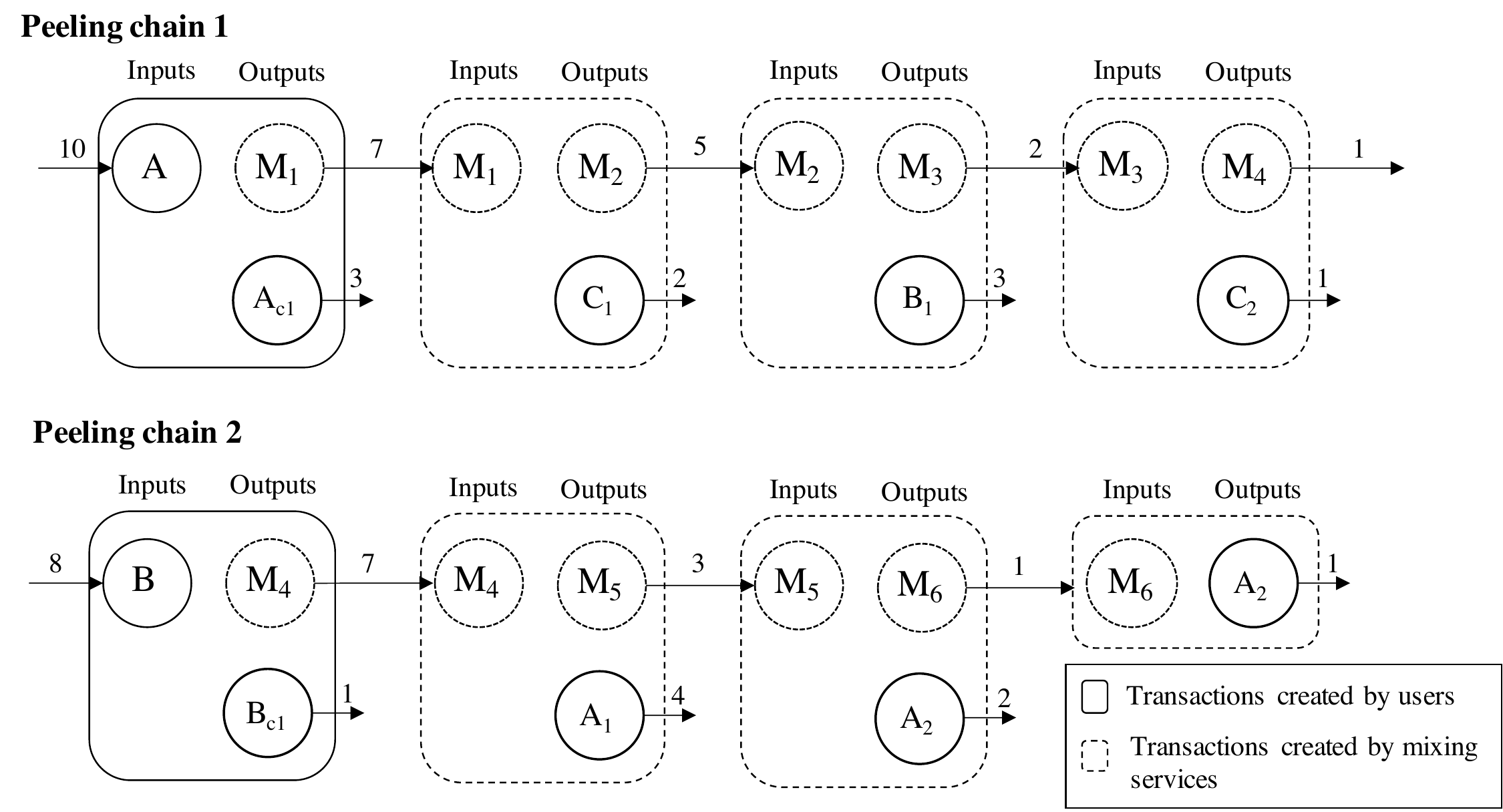}
\caption{{\small An example of the swapping mechanism. In this figure, we use $M_1$ to $M_6$ to denote addresses maintained by the mixing service.
By swapping different user inputs and outputs, the relationship anonymity for all addresses is preserved. For instance, the relationship from $A$ to $A_1$ and $A_2$ is anonymized. 
}}
\label{fig:abs_ex_swapping}

\end{figure}

\subsubsection{Type I -- $\mathcal{M}_S$}
\label{subsubsec:swaapingmechanism}
The basic idea of $\mathcal{M}_S$ is 
to swap the inputs and outputs from different users
to preserve relationship anonymity.
Note that in any $T_{\mathcal{M}}$ of $\mathcal{M}_S$, there is only one user output.

Figure~\ref{fig:abs_ex_swapping} gives an example.
Instead of directly sending 7 BTC from $A$ to $A_1$ and $A_2$,
the mixing service
will swap the outputs of $B$ to them.
Similarly, $B_1$ is swapped from outputs of $M_2$, which originates from $A$.

Despite the simple and effective idea of swapping, there is an important assumption that $T_{\mathcal{M}}$s are hidden by the service.
Otherwise if we can identify all $T_{\mathcal{M}}$s, the original relationships between inputs and outputs can be recovered.
For instance, if we discover all $T_{\mathcal{M}}$s in Figure~\ref{fig:abs_ex_swapping}, 
then we can find out that the output value $M_1$ is equal to the input value $M_4$ of a mixing transaction.
Consequently, we can infer that $M_1$ and $M_4$ are swapped and the original output of $A$ is $A_1$ and $A_2$.

To prevent $T_{\mathcal{M}}$s from being identified, the concept of \textit{peeling chain} was observed in the wild~\cite{moser2013mixing}.
A peeling chain is a set of transactions generated by mixing services that form a chain to distribute outputs.
The unique property of the peeling chain is that
transactions in the chain are similar to normal user transactions with two outputs~\cite{harrigan2016clustering}. 
Thus, $T_{\mathcal{M}}$s cannot be easily distinguished from normal user transactions.

For a $T_{\mathcal{M}}$ in the peeling chain, one of the outputs is used to generate the output for the specified output address and another is used for the change, which in turn becomes the input of the next chain node.
In Figure~\ref{fig:abs_ex_swapping},
the input to $M_1$ is separated into two outputs, one is $2$ BTC to $C_1$
and another is $5$ BTC to $M_2$.
The latter output then becomes the input to $T^{\prime}_{\mathcal{M}}$.
The peeling chain will be detailed in Section~\ref{subsec:graph_analysis}.

\begin{figure}[t]
    \centering
    \includegraphics[width=0.5\textwidth]{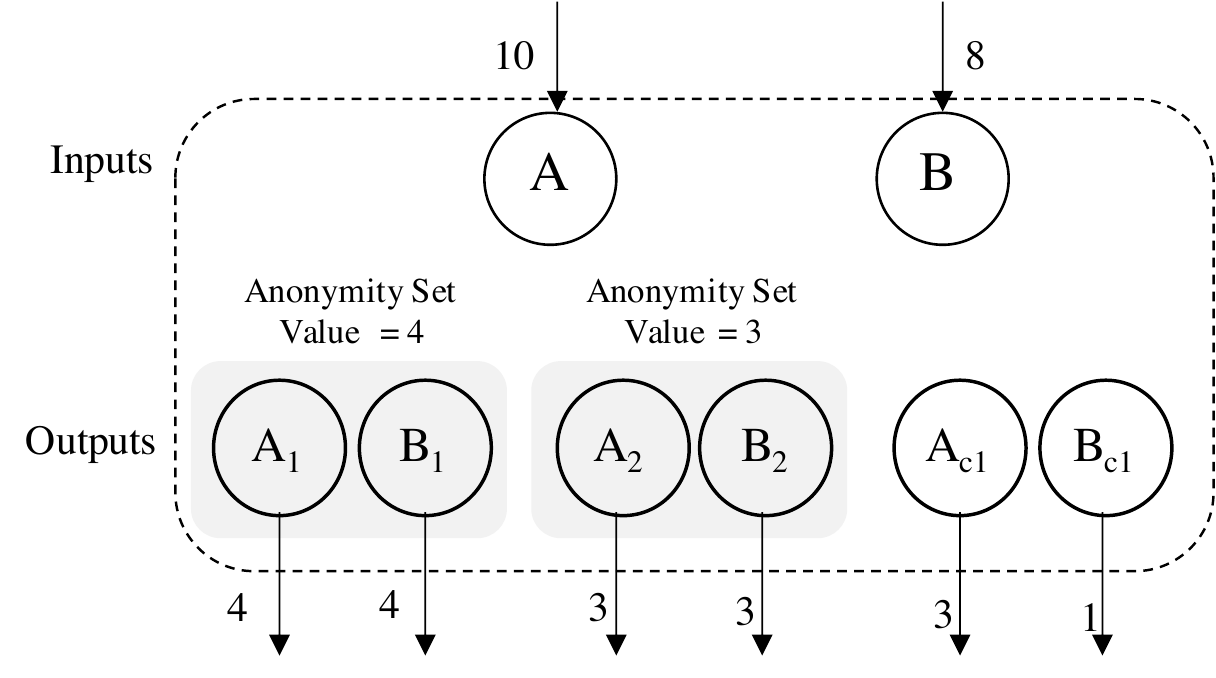}
    \caption{{\small
    An example of obfuscating with a single mixing transaction.
    The mixing service generates two anonymity sets with the size 4 and 3, respectively.
    }}
    \label{fig:abs_ex_coinjoin}
    \end{figure}
    
    \begin{figure}[t]
    \centering
    \includegraphics[width=.5\textwidth]{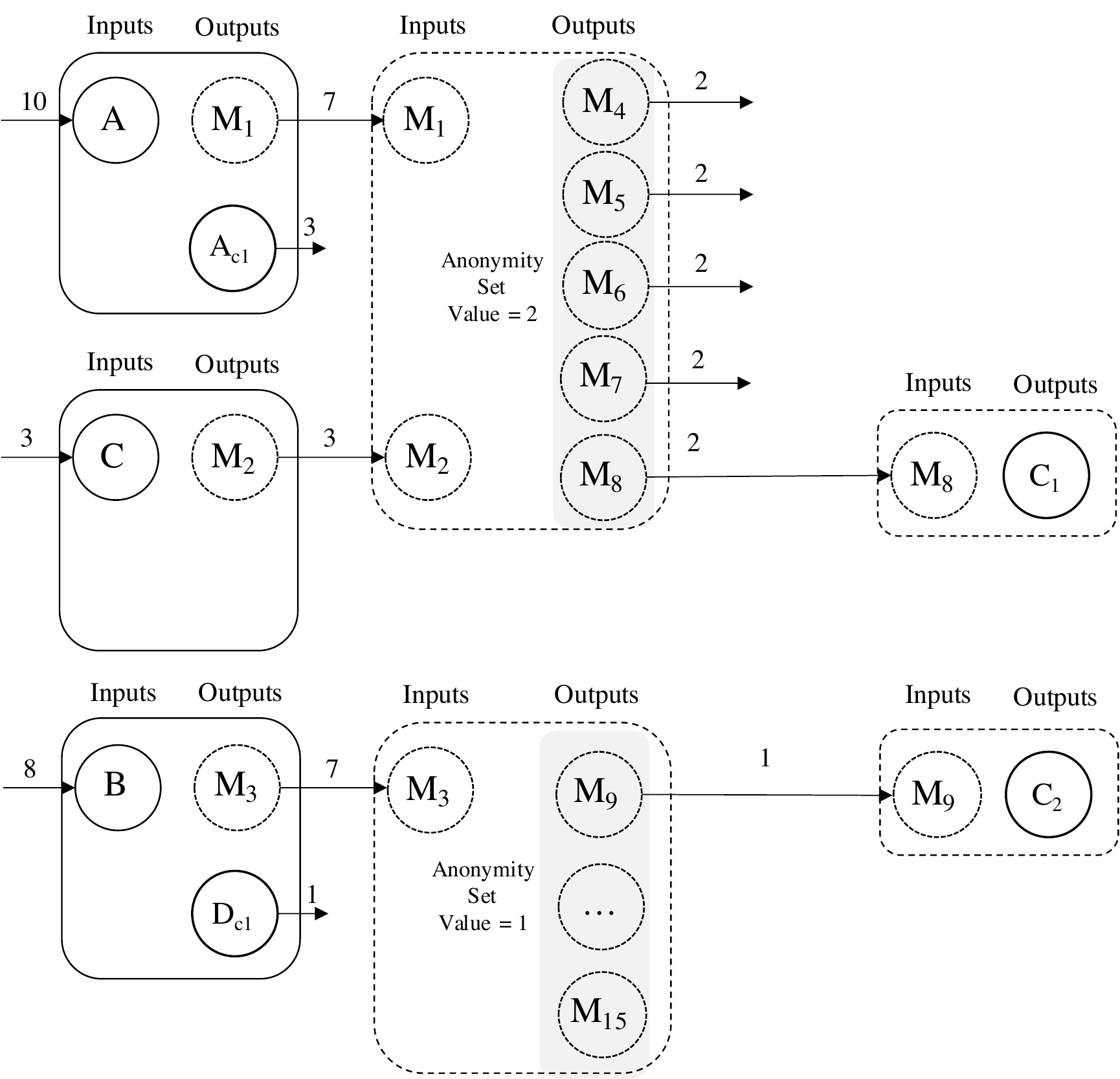}
    \caption{{\small
    An example of obfuscating with multiple mixing transactions.
    The mixing service generates multiple anonymity sets with different
    values ($2$ and $1$ in this case). From the figure, $C_1$ and $C_2$ are
    from $M_8$ and $M_9$ with input value $2$ and $1$ BTC. Other outputs, e.g., $M_4$ and $M_{15}$ will be used to mix other inputs.
    }}
    \label{fig:abs_ex_anonset}
    \end{figure}

\subsubsection{Type II -- $\mathcal{M}_O$}
\label{subsubsec:obfuscatingmechanism}
$\mathcal{M}_O$ aims to preserve the relationship anonymity
by breaking the matching procedure between user inputs and outputs.
It is achieved by using anonymity sets to hide user outputs. 
Note that in any $T_{\mathcal{M}}$ of $\mathcal{M}_O$, there is at least one anonymity set $A_N$.

Figure~\ref{fig:abs_ex_coinjoin} shows an example of a mixing transaction of $\mathcal{M}_O$.
There are two $A_N$s, and the outputs in each of them are indistinguishable.
For example, it is impossible to determine which output $A_1$ or $B_1$ in the first group originates from input $A$.
It is hard to identify the real outputs for each input without additional information.

Moreover, the obfuscating procedure could be achieved within single or multiple
$T_{\mathcal{M}}$s. 
Figure~\ref{fig:abs_ex_coinjoin} and Figure~\ref{fig:abs_ex_anonset} are two examples, respectively. 
Compared with a single $T_{\mathcal{M}}$, multiple $T_{\mathcal{M}}$s can generate fine-grained outputs with user-specified values. 
For instance, when using the mixing service, users $C$ can specify that the outputs for $C_1$ and $C_2$ are $2$ and $1$ BTC.
Conversely, the service determines the value for each $A_N$ for each transaction in the case of
the single $T_{\mathcal{M}}$.
Note that in both cases, there are some transactions that involve multiple inputs and outputs with the same value.

\section{Methodology}
\label{sec:method}

In this section, we will introduce our methodology to
analyze mixing services. We will first select representative
mixing services and then collect sample transactions.
After that, based on these transactions, we will perform 
transaction-based analysis to identify mixing mechanisms
used by these services.
 
\subsection{Select Representative Mixing Services}
\label{subsec:selection_method}
To select representative mixing services,
we use \textit{BitcoinTalk}~\cite{bitcointalk} and other public media as the information sources.
As the biggest Bitcoin-dedicated forum, BitcoinTalk has served as the official forum for Bitcoin.
Besides, we also pay attention to reports from other public media reports.
For example, ShapeShift was investigated and reported by the Wall Street Journal~\cite{wsjshapeshift} for being used for money laundering.

\subsection{Collect Sample Transactions}
\label{subsec:acquisition_method}
To analyze a mixing service, we first obtain the transactions that
are used for the mixing purpose.
We denote them as \textit{sample transactions} in our study.
For a typical mixing service, sample transactions include \textit{input transactions} by users to send inputs to the service, and \textit{output transactions} by the service to send mixed outputs to user specified output addresses.
The input transactions are initiated by users, while the mixing and output transactions are initiated by the service.

Specifically, the following two complementary methods are used.

\noindent\textbf{Method I: Interacting with Mixing Services.}
We use the mixing service by sending Bitcoins to the service
and then collecting the input and output transactions.
This method would be restricted by the budget constraint as
some mixing services may place a high input threshold or may charge a high mixing fee.
Therefore, we only conduct a small number of experiments using this method.

\noindent\textbf{Method II: Using Public APIs.}
Some mixing services provide public APIs to facilitate their usage.
For instance, they provide APIs for users to query detailed information
and update status of a mix or inspect statistics of a mixing service.
Fortunately, the returned data usually contains redundant information that
help to reveal or reconstruct users' mixing records.

\subsection{Basic Transaction Analysis}
\label{subsec:graph_analysis_method}

Based on sample transactions, our next step is to determine the
mixing mechanism used by the service and understand its mixing process.
This is achieved by performing a transaction-based analysis on sample transactions.
There are two challenges in performing such analysis.
In the following, we will discuss these challenges and our methods to solve them.

\subsubsection{Challenges}
\label{subsubsec:bchallenges}
We have to deal with the following two challenges in our analysis.

\noindent
\textbf{Challenge I: Identify Address Types.}\tab
When constructing the transaction graph from sample transactions,
we first need to distinguish the users' addresses and those addresses used by mixing services. 
Otherwise, our graph would be too big (with too many nodes and edges) to be analyzed and introduces false positives.

To address this challenge, we pay special attention to the user behavior
in transaction analysis, and observe that addresses used by the mixing services
tend to belong to the same type (address types are introduced in Section~\ref{subsec:bgbitcoin}).

Based on this observation, we can distinguish addresses when there
exist two different types of addresses in a sample transaction.
For example, if there are two types of addresses in a transaction and one of them is determined to be used by mixing services, then addresses of the other type are considered to be used by users.
We can then prune the transaction graph to remove users' addresses and transactions.

\noindent
\textbf{Challenge II: Identify Peeling Chains.}\tab
Although peeling chains are commonly
observed~\cite{moser2013mixing, de2017analysis}, they have not been carefully analyzed.
A peeling chain can be modeled as a structure consisting of three components, including \textit{starting point}, \textit{chain nodes} and \textit{ending point}, which will be analyzed and distinguished accordingly.

Specifically, a \textit{starting point} is the transaction that a user sends the input to an address given by the mixing service, e.g., $M_1$ in 
Figure~\ref{fig:abs_ex_swapping}.
There are two possible methods to distinguish the starting point in
sample transactions.
First of all, based on the \textit{multi-input and change address heuristic} ~\cite{reid2013analysis}, this transaction initiated by users should have only two outputs, one of which is the service provided deposit address and another is the change address.
Secondly, the address type can also be used to distinguish the change output from
the service output, if the two outputs have different address types.

\textit{Chain nodes} are used to distribute user outputs and continue the peeling chain.
The structure of chain nodes is simple with one input (a reference to output from the previous node), and two outputs (one for user output and another for the successive node).
However, there exist cases that chain nodes are indistinguishable from the starting point.
In this case we trace backwards until a transaction with multiple inputs are found, and manually inspect them to find the starting point.

An \textit{ending point} is the end of a peeling chain. The remaining changes at the tail will be handled by the service. For instance, these changes could be used as inputs for other mixes.
Our observation suggests that if the changes from a chain node is used in a transaction with many inputs, the corresponding chain node can be regarded as the ending point.
Mixing services will collect these remaining changes for future use.

\subsubsection{Determine Mixing Mechanisms}
\label{subsubsec:determine_mechanism_method}
During the analysis, we define the \textit{context} as the destinations
of inputs and sources of outputs in sample transactions. 
We determine the mixing mechanism by examining contexts of these transactions.

As introduced in Section~\ref{sec:abs}, the major transaction-level difference of the swapping and the obfuscating mechanism comes
from the \textit{output pattern}. For the swapping mechanism,
mixing outputs are consecutive, while the outputs are centralized for the obfuscating mechanism.
We examine the context of outputs in sample transactions, and use the difference
as a criteria to distinguish the mixing mechanism. We use the following
heuristics in this study.

\begin{itemize}
	\item If most transactions have two outputs, and they form a chain using change addresses in the context of each outputs, then the service uses the swapping mechanism.
	\item If there is a transaction generating outputs with identical values (i.e., anonymity sets) in the context of each output in sample transactions, then the service uses the obfuscating mechanism.
\end{itemize}

\subsubsection{Understand Mixing Process}
\label{subsubsec:analyze_workflow_method}
After the mixing mechanism is determined, we then
figure out the mixing process, i.e., how the service 
performs the mixing.

\noindent\textbf{Swapping Mechanism.}
For mixing services using the swapping mechanism,
the peeling chain is the central structure used in a mixing process.
In our study, we first draw the transaction graph
and then identify the components of a peeling chain
leveraging the previously discussed definitions in Section~\ref{subsubsec:bchallenges}.

\noindent\textbf{Obfuscating Mechanism.}
For mixing services using the obfuscating mechanism, we focus on transactions that generate 
anonymity sets. Specifically, for each input and output in the sample transaction,
we find corresponding transactions that generate anonymity sets to spend
the input or send the output.
\begin{algorithm}[t]
	\SetAlgoLined
	\SetKwData{True}{true}
	\KwData{Seed transaction set $S$ from the mixing service.}
	\KwResult{Expanded transaction set $E$, in which each element is highly likely to be related to the mixing service.}
	Initialize a queue $\mathbf{Q}$ with all element in $S$\;
	Initialize $E$ to be an empty set\;
	\While{Queue $\mathbf{Q}$ is not empty}{
		Take a transaction $T$ from $\mathbf{Q}$\;
		Put $T$ into the set $E$\;
		\For{every output $O$ in $T$} {
			Find transaction $T_O$ that uses the output $O$\;
			\For{every input $I$ in $T_O$} {
				Find transaction $T_I$ referred by input $I$\;
				\If{$T_I$ generates anonymity sets \textbf{and} $T_I$ not in $E$} {
					En-queue $T_I$ into $\mathbf{Q}$\;
				}
			}
		}
	}
	\caption{The Seed-Expansion Algorithm}
	\label{algo}
\end{algorithm}
 
\subsection{Advanced Transaction Analysis}
\label{subsec:behavioral_analysis_method}

Besides the previous analysis,
we also conduct further analysis to identify mixing transactions
for services using the obfuscating mechanism.
This is important as it helps to inspect money flow to mixing services and
investigate money laundering activities.
We take a two-step analysis with seed inputs.

\noindent\textbf{Step I: Identify Anonymity Sets.}
Our first step is to identify anonymity sets using seed inputs,
which are fed into the mixing service with service-provided
addresses (e.g., $M_1$ in Figure~\ref{fig:abs_ex_anonset}). Then, 
we can locate addresses in the anonymity set by finding outputs with the same value.
We color each address ($M_4$ to $M_8$) in the identified set.
We also color the outputs for transactions that take the colored
address as inputs, e.g., the address $C_1$ is colored.

\noindent\textbf{Step II: Identify More Anonymity Sets.}
We then perform further analysis to identify more transactions.
In particular, if we find a transaction with multiple inputs
that takes a colored address as input, then we color other input
addresses. For example, if we find there exists a transaction
with $C_1$ and $C_2$ as inputs, then we will color $C_2$ too. 
We do not color $C_2$ in the previous step since we only use
the input $A$ as the seed input. Input to $B$ is not the seed input.

After that we perform a backward analysis from $C_2$. In particular,
we move backward from address $C_2$ and try to find
transactions that have the same output values, e.g., from $M_9$ to $M_{15}$.
These outputs with the same value means that new anonymity sets are detected. We color them
and perform the similar analysis from each address.

During this step, we may not find any anonymity set.
In this case, we will remove the color accordingly.
For instance, if $E_1$ and $C_1$ are inputs for a transaction, then $E_1$
will be colored. However, $E_1$ may come from outputs of normal user transactions.
In this case, we will remove the color for $E_1$.

In summary, the whole analysis algorithm is shown in Algorithm~\ref{algo}.
By applying it with seed inputs, we can identify mixing transactions and
corresponding addresses used for mixing.

\section{Evaluation Results}
\label{sec:results}
In this section, we apply the proposed methodology
in Section~\ref{sec:method} and summarize the results.

\subsection{Selected Services}
\label{subsec:selection_result}
In this paper, we select the following four services for evaluation.

\noindent\textbf{Chipmixer}~\cite{chipmixer} 
is one of the most popular mixing services. 
Its popularity originates from its ``Pay What You Want'' (PWYW) pricing strategy.
In addition, it was reported that Chipmixer was used by the attacker to launder over 4,000 BTC~\cite{clainreport}.
 
\noindent\textbf{Wasabi Wallet}~\cite{wasabiwallet}
is one of the officially recommended desktop Bitcoin wallets~\cite{choosewallet}, and the only (currently available and popular) wallet with the built-in CoinJoin functionality~\cite{wasabiwallet}.
 
\noindent\textbf{ShapeShift}~\cite{shapeshift} 
is one of the most famous cryptocurrency converters. According to the report of the Wall Street Journal, it was used as a money laundering tool for over 9 million dollars of tainted funds over a time period of two years. Due to the pressure from the public media and regulators, it has applied Know-Your-Customer (KYC) policy and requires personal identification to set up an account.
 
\noindent\textbf{Bitmix.biz}~\cite{bitmixbiz} 
was announced in August, 2017~\cite{bitmixbizann}.
It claimed to have some improvements over its predecessors like dust-attack prevention, letters of guarantee (to redeem funds on exceptions), and randomized transaction fees and delays. 
The wider range of supported cryptocurrencies (Bitcoin, Litecoin and DASH) and lower mixing fee (from 0.4\%) also contributes to its popularity.

\subsection{Sample Transactions Collection}
\label{subsec:acquisition_result}
As introduced in Section~\ref{subsec:acquisition_method}, there are two complementary
methods to obtain sample transactions.
We first conduct a complete analysis on these services to determine which method to use.
For Wasabi Wallet and ShapeShift, we find public APIs that can be used to obtain sample transactions.
In contrast, for Chipmixer and Bitmix.biz we resort to interaction with the service.
Table~\ref{tab:sample_tx} summarizes the collected sample transactions.

\begin{table}[t]
    \caption{Sample transactions obtained for selected services.}
    \label{tab:sample_tx}
    \begin{center}
    \scalebox{1.0}{
    \setlength{\tabcolsep}{1mm}{
        \begin{tabular}{cccc}
        \toprule
        \multirow{2}*{\textbf{Service}}    & \multirow{2}*{\textbf{Method}}    & \textbf{\# of Samples}        \\
        ~ & ~   & \textbf{Obtained}     \\
        \midrule
        Chipmixer           & Interacting with the Service      &  $20$ ($5$ inputs + $15$ outputs)    \\
        Wasabi Wallet       & Using Public APIs             & $4,850$                             \\
        ShapeShift          & Using Public APIs             & $6,381$ (Bitcoin) + $1,089$ (Litecoin)  \\ 
        Bitmix.biz          & Interacting with the Service      & $20$ ($6$ inputs + $14$ outputs)    \\
        \bottomrule
        \end{tabular}
    }}
    \end{center}
\end{table}

\subsubsection{Interacting with Services}
\label{subsubsec:results_interacting}
In the following, we will describe the details of obtaining sample transactions by interacting with Chipmixer and Bitmix.biz, respectively. We performed
the collection from October, 2019 to February, 2020.

\noindent
{\textbf{Chipmixer.}}\tab
According to its pricing strategy, Chipmixer can be used as a free service. However, it only recognizes inputs up to 3 digits after the decimal point and any trailing value will be considered as service fees or donations.

This service first provides a generated address for users to send inputs. 
When an input is confirmed, the next step is to decide how to distribute the input into \textit{chips}~\footnote{Chips are defined as user outputs with predefined values~\cite{chipmixer}.}. 
After the distribution of chips, users can withdraw these chips by either importing the provided private keys or specifying output addresses separately.

In total, we conducted 5 experiments and received 15 outputs.

\noindent
{\textbf{Bitmix.biz.}}\tab
Users of Bitmix.biz can directly set mixing parameters and send mixing requests.
Parameters include output addresses, the delay from the mixing request to output received, value distribution (distributions for each address) and overall transaction fees.
After receiving a request, the service will provide a temporary address to receive inputs.
Once the inputs are confirmed, it will send corresponding outputs according to the requested delay.

In total, we conducted 6 experiments and received 14 outputs.
 
\subsubsection{Using Public APIs from Services}
As stated in Section~\ref{subsec:acquisition_method},
services may provide public APIs used to obtain sample transactions.

\noindent
{\textbf{Wasabi Wallet.}}\tab
It provides two APIs to fetch mixing-related data: 1) the API \texttt{states}~\cite{wasabiapistates} is used for the clients to query and update current phase and status of current CoinJoin transaction; and 2) the API \texttt{unconfirmed}~\cite{wasabiapiunconfirmed} broadcasts transaction hashes of all successful CoinJoin transactions before they are confirmed.

These two API are for status querying and updating purposes. However,
the Wasabi Wallet server does not require any authentication to access them.
Therefore, we used a crawler to periodically retrieve information.
The crawler accessed these APIs every 1 minute and continued
for 82 days (from December 26, 2019 to March 15, 2020).

In total, we gathered $4,850$ transactions. We will use these transactions as
the seed set for our
experiment.

\noindent
{\textbf{ShapeShift.}}\tab
There are two key APIs that can be used to obtain sample transactions. The first API is called \texttt{recenttx}~\cite{ssapirecenttx}. It provides information about all recent convert records in ShapeShift. Each convert record is represented by a tuple of \texttt{<curIn, curOut, timestamp, value>}, which represents the cryptocurrency type of input and output, timestamp of the convert, and input currency value in decimal.
The second API is called \texttt{txstat}~\cite{ssapitxstat}.
For a given address, it provides detailed information if the address is used by ShapeShift.
While ShapeShift requires a registered account and personal identification information, using these APIs requires no authentication.

In total, we crawled $27,411$ convert records from December 11, 2019 to March 18, 2020. We focused on converting records from Bitcoin to other cryptocurrencies. In the crawled records, we found $7,067$ records with Bitcoin as the input cryptocurrency.

To further identify corresponding transactions for a given convert record, we propose a refined algorithm based on~\cite{yousaf2019shapeshift}.
This algorithm consists of three steps.
First, we obtain a list of recent cryptocurrency convert records using the \texttt{txstat} API.
After that, for each record (with value $v$ and timestamp $ts$), we locate candidate transactions with the closest values to $v$ and closest timestamps to $ts$. 
Finally, these transactions will be further validated by applying the \texttt{txstat} API.
We have applied this algorithm on crawled $7,067$ records, and successfully matched $6,381$ convert records ($90.29$\% of all records) with detailed information.

So far, the transactions we obtained are input samples, we also need output samples to analyze the complete convert workflow.
Besides, we want to analyze where output Bitcoin comes from in the case of converting other cryptocurrencies to Bitcoin.
To this end, we chose Litecoin by its popularity in ShapeShift, and found $1,097$ records converting Litecoin to Bitcoin.
Then we apply the proposed algorithm to these records and $1,089$ ($99.27$\%) records are matched with detailed information.

\subsection{Basic Transaction Analysis}
\label{subsec:graph_analysis}
We have applied basic transaction analysis discussed in Section~\ref{subsec:graph_analysis_method} to the four selected services.
In the following, we briefly describe the results and findings in our analysis.

\begin{figure}[t]
	\centering
	\includegraphics[width=.9\textwidth]{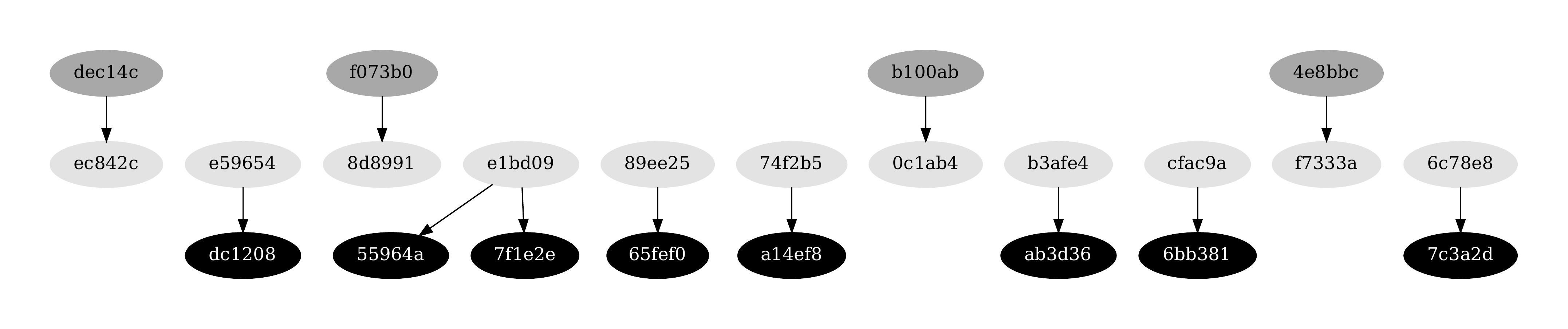}
	\caption{The transaction graph of the Chipmixer experiment.
		Gray and black nodes are our input and output transactions, respectively.
		Transactions in light gray nodes generate anonymity sets.
	} 
	\label{fig:cm_experiment}
\end{figure}

\subsubsection{Determining Mixing Mechanisms}
Obviously, as Wasabi Wallet implements the CoinJoin protocol~\cite{wasabiwallet} that generates anonymity sets, it uses the obfuscating mechanism.
Apart from the Wasabi Wallet, the mixing mechanisms used by other three services are determined by analyzed the transaction graph of the obtained transactions.

To determine the mixing mechanism used by Chipmixer, we first plot a transaction graph of sample transactions and their contexts.
Figure~\ref{fig:cm_experiment} is the transaction graph for the experiments conducted with Chipmixer.
The figure shows that all of our inputs (gray nodes) are immediately spent by mixing transactions (light gray nodes) by the service, and our outputs (black nodes) also come directly from them.
Mixing transactions in light gray nodes generate anonymity sets, indicating that Chipmixer uses the obfuscating mechanism.
Because all outputs from these mixing transactions are of specified value (as mentioned in Section~\ref{subsec:acquisition_method}), Chipmixer generates a fixed number of large anonymity sets.

\begin{table}[t]
    \caption{Mixing mechanisms used by services.}
    \label{tab:mixing_mechanisms}
    \begin{center}
    \scalebox{0.9}{
    \setlength{\tabcolsep}{1mm}{
        \begin{tabular}{rcc}
        \toprule
        \textbf{Service}    & \textbf{Swapping mechanism}       & \textbf{Obfuscating mechanism}\\
        \midrule
        Chipmixer           & ~                                 & $\surd$  \\
        Wasabi Wallet       & ~                                 & $\surd$  \\
        ShapeShift          & $\surd$                           & ~\\ 
        Bitmix.biz          & $\surd$                           & ~\\
        \bottomrule
        \end{tabular}
    }}
    \end{center}
\end{table}

\begin{figure}[t]
\centering
\includegraphics[width=.9\textwidth]{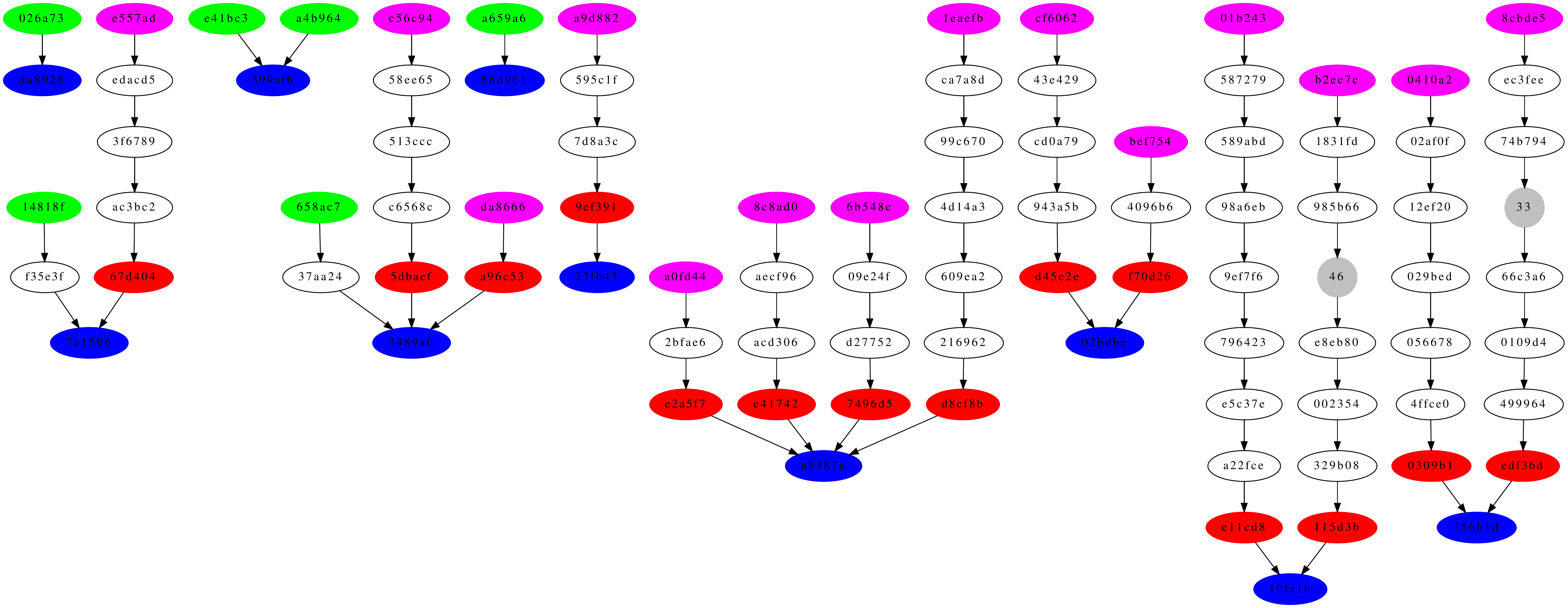}
\caption{The transaction graph of Bitmix.biz experiments.
Green nodes represent our input and output transactions.
Blue nodes are ending points, and magenta nodes are potential starting points of the peeling chains.
Gray circles with numbers denote omitted long chains.
User output in each chain node is also omitted.
}
\label{fig:bitmix}
\end{figure}

Similarly, we applied the same approach to the other two services.
Based on the corresponding transaction graph,
we conclude that they use the swapping mechanisms.
For example, in Figure~\ref{fig:bitmix}, all of our outputs come from mixing transactions with only two outputs (i.e., no anonymity sets get involved).
Tracing our outputs backward shows several chains, in which most transactions have single input and two outputs. They are connected with change addresses.
Again, according to Section~\ref{sec:abs}, this is a feature of the peeling chain.
Results of all these services are summarized in Table~\ref{tab:mixing_mechanisms}.

\subsubsection{Understanding Mixing Process}
To better understand the mixing services, we need to figure out their mixing workflows.

\begin{figure}[t]
	\centering
    \subfigure[Chipmixer.]{
        \includegraphics[width=.45\textwidth]{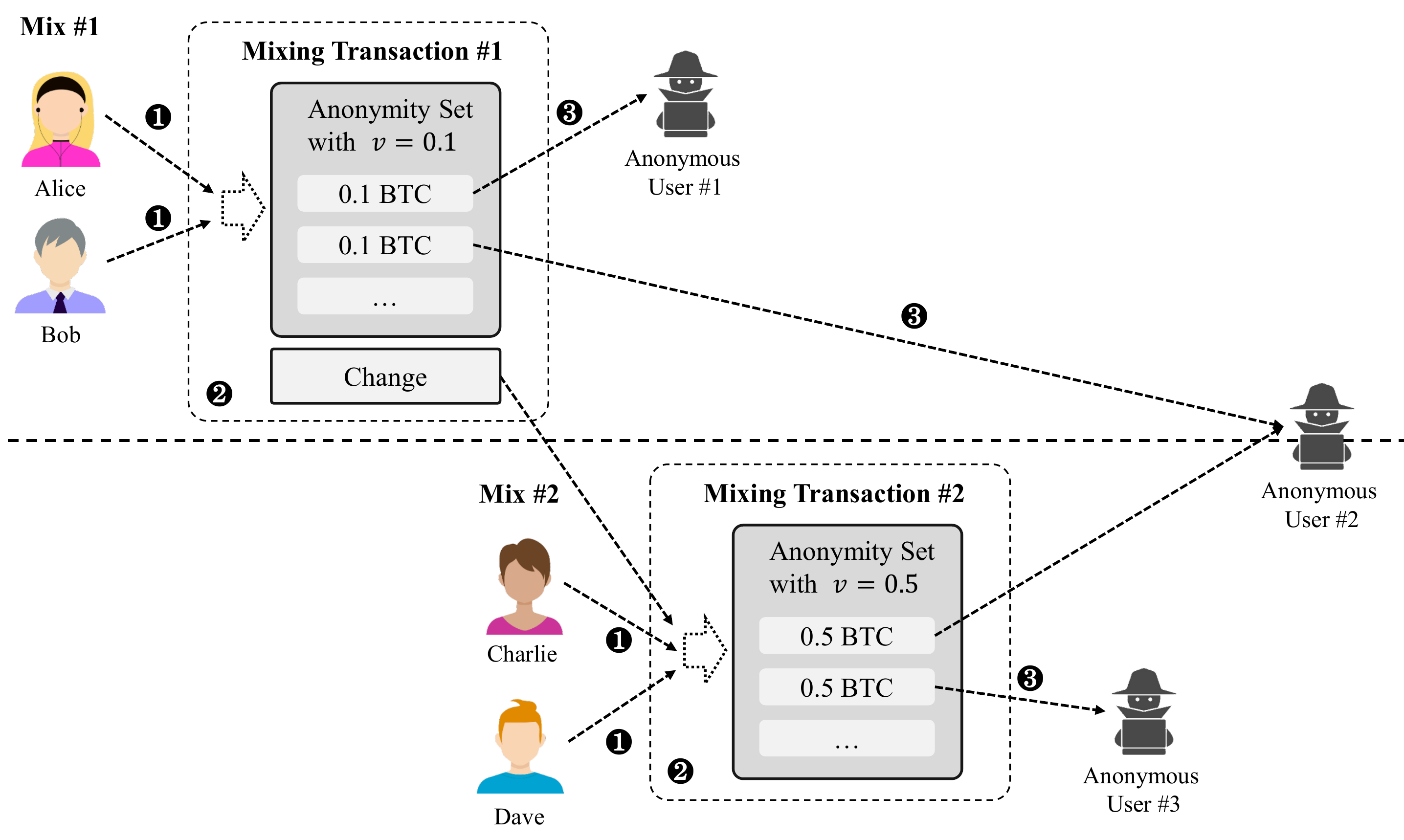}\label{fig:workflow_chipmixer}
    } \hfill
    \subfigure[Wasabi Wallet.]{
        \includegraphics[width=.45\textwidth]{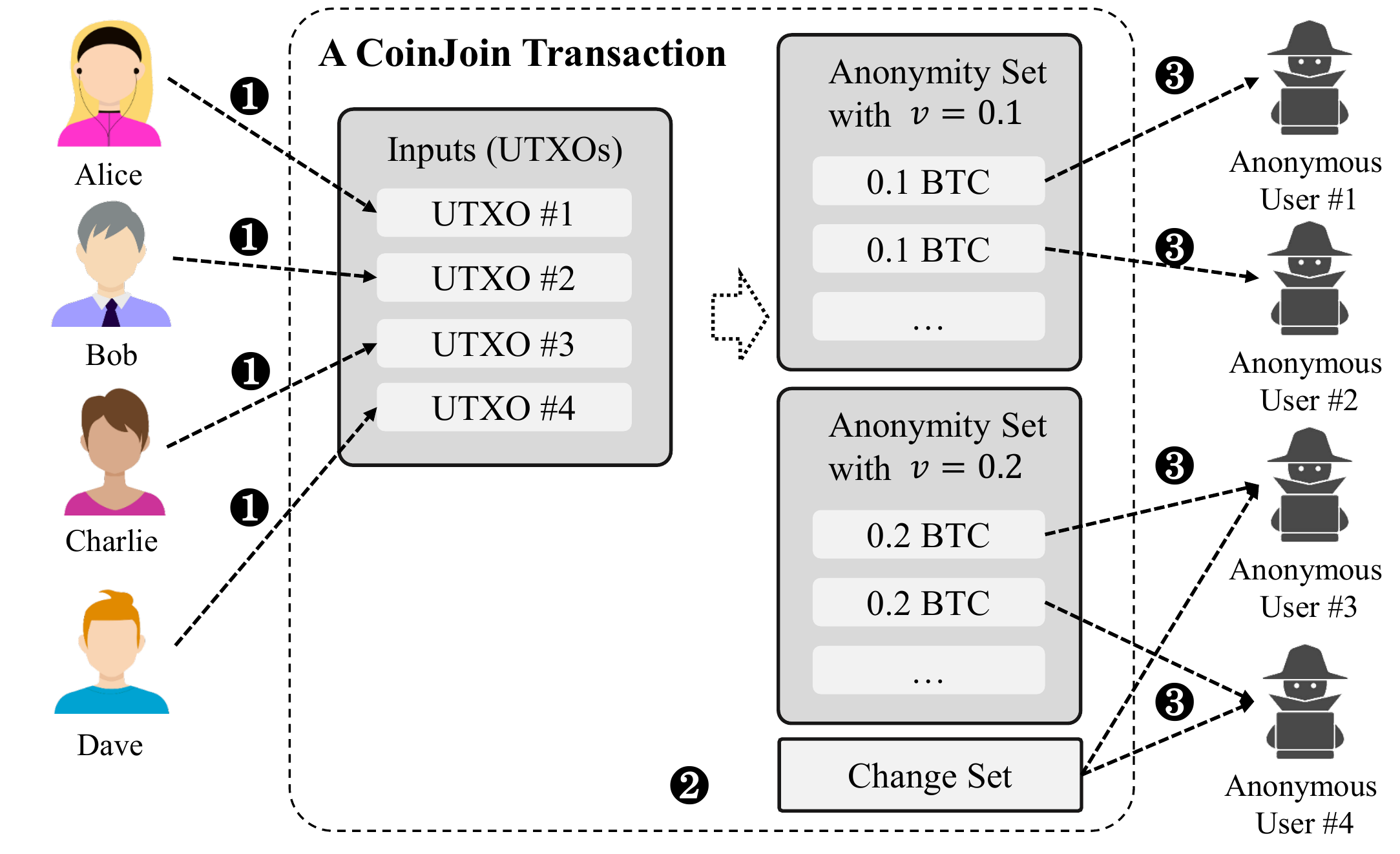}\label{fig:workflow_wasabi}
    } 
\caption{Examples for the mixing process of Chipmixer and Wasabi Wallet.}
\label{fig:workflows_chipmixer_wasabi}
\end{figure}

\noindent
{\textbf{Chipmixer.}}\tab
Users first send their inputs to the service. Then, the service generates chips (in anonymity sets) using mixing transactions. Lastly, the service sends those chips back to users.

Figure~\ref{fig:workflow_chipmixer} gives an example with two mixes. In mix \#1, two users (Alice and Bob) send their inputs to the service. These inputs are aggregated by the service into mixing transaction \#1, which generates an anonymity set with value $0.1$. The outputs in this anonymity set will be distributed to users. If the inputs do not fit the anonymity set properly, then there will be change left as an input for another mix. For example, the inputs of mix \#2 come from another two users (Charlie and Dave) along with the change of mix \#1. The anonymity set generated by mix \#2 have a value of $0.5$, which fits the inputs without any change. 

\noindent
{\textbf{Wasabi Wallet.}}\tab
Unlike other services that create addresses for users to deposit Bitcoin, this service requires users to send UTXOs and output addresses in the wallet.
Then, the service creates a number of anonymity sets with a change set in one CoinJoin transaction. Finally, the service transfers outputs to corresponding addresses. 

Figure~\ref{fig:workflow_wasabi} gives an example. In step 1, users of this CoinJoin round (i.e. this mix), Alice, Bob, Charlie and Dave, submit UTXOs they want to mix and output addresses to the service.
Then in step 2, two anonymity sets with value $0.1$ and $0.2$ are generated. Finally in step 3, outputs in anonymity sets and changes are sent correspondingly to the output addresses. As a result, outputs in the anonymity set are hidden, but the changes are not anonymized (not in an anonymity set) and require further CoinJoin rounds.

\noindent
{\textbf{ShapeShift.}}\tab
Users first send their Bitcoin to addresses provided by the service and specify output addresses in the other blockchain network. Then the service takes responsibility for the mixing by performing cross-blockchain transactions. Finally, users can receive coins from the other blockchain.

\begin{figure}[t]
	\centering
	\subfigure[ShapeShift.]{
		\includegraphics[width=.45\textwidth]{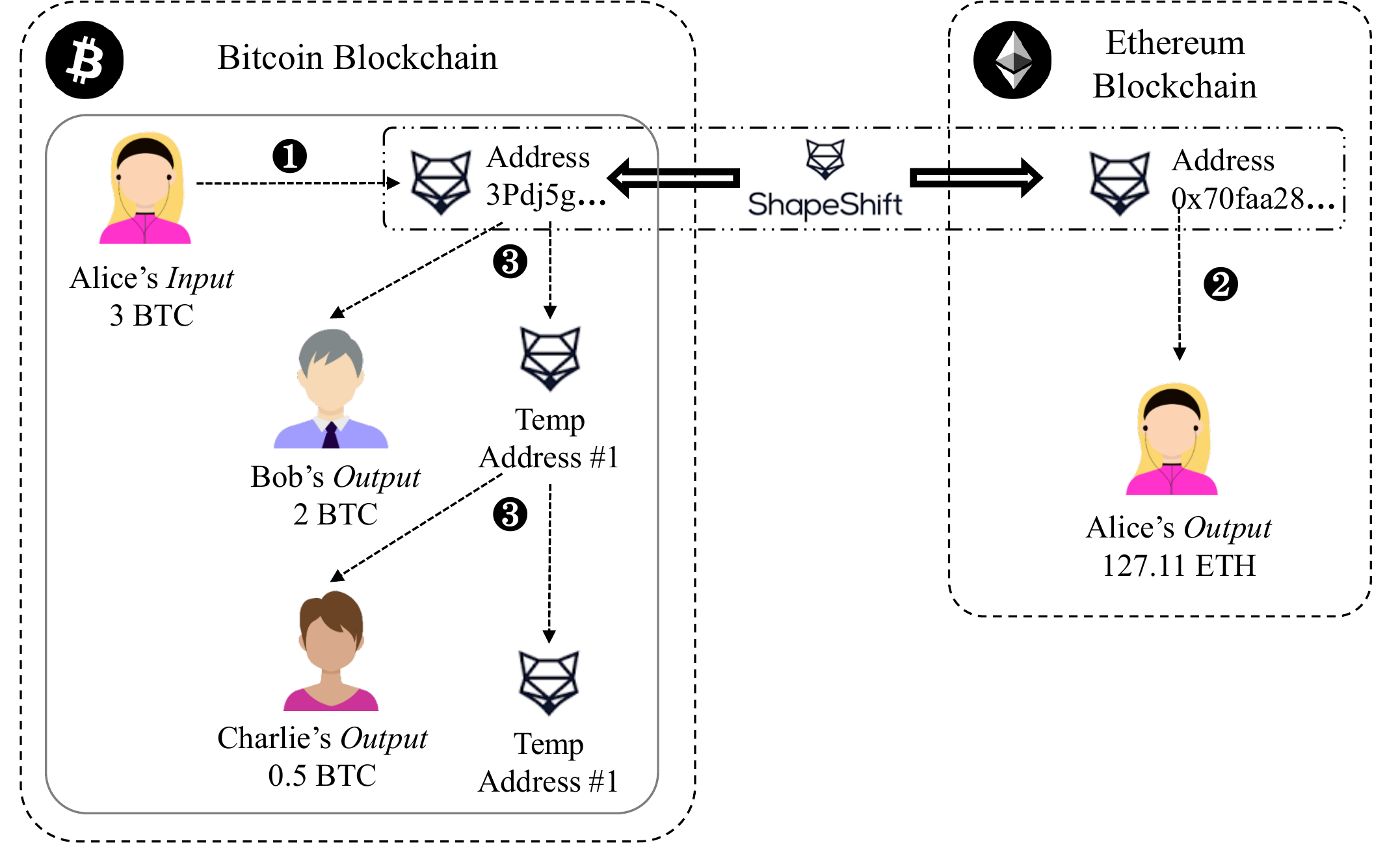}\label{fig:workflow_shapeshift}
	} \hfill 
	\subfigure[Bitmix.biz.]{
		\includegraphics[width=.45\textwidth]{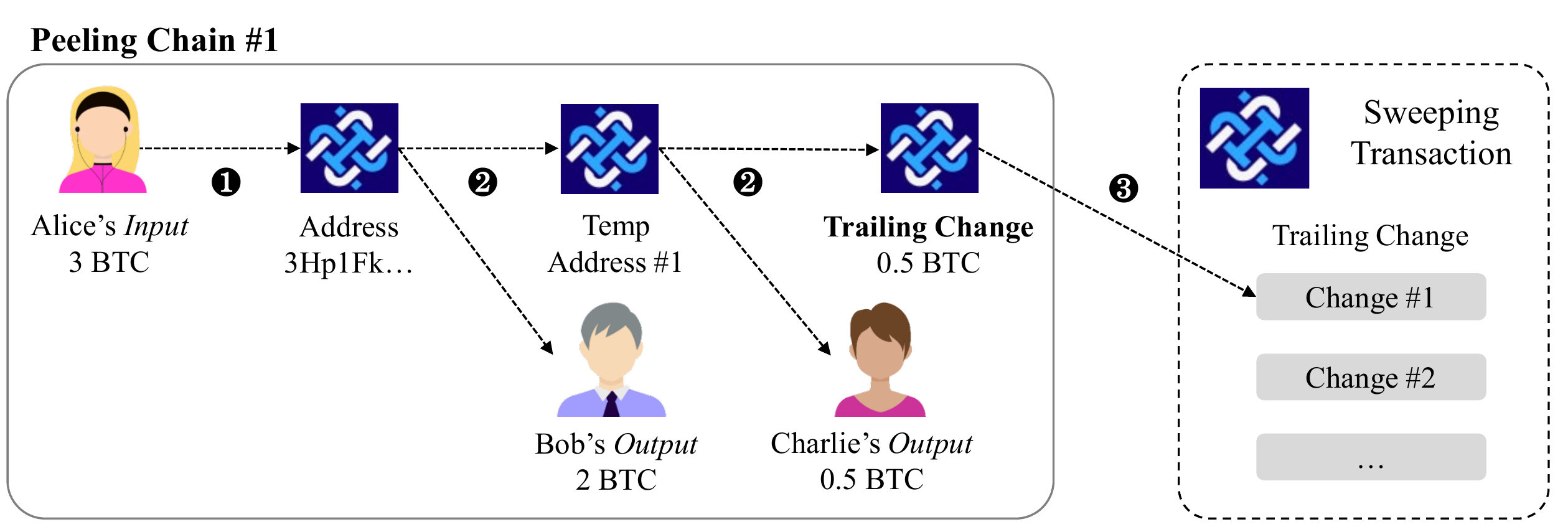}\label{fig:workflow_bitmix}
	} 

	\caption{Examples for mixing workflows of ShapeShift and Bitmix.biz.}
	\label{fig:workflows_shapeshift_bitmix}
\end{figure}

Figure~\ref{fig:workflow_shapeshift} gives a concrete example. In the Bitcoin network, Alice sends $3$ BTC to ShapeShift and receives $127.11$ Ether in Ethereum later. Obviously, this service has to make efforts (e.g. in collaboration with cryptocurrency exchanges) to break even among different blockchain platforms.
Due to the swapping mechanism, the Bitcoin sent by Alice will be organized as a peeling chain to distribute Bitcoins to other users (e.g., Bob and Charlie in this figure) who swap other cryptocurrencies for Bitcoin.

\noindent
{\textbf{Bitmix.biz.}}\tab
Users first send their Bitcoin to addresses provided by the service. Then, the service creates peeling chains to distribute the outputs. Finally, users receive their outputs from the chain nodes in the peeling chain. 

An example of a peeling chain for Bitmix.biz is shown in Figure~\ref{fig:workflow_bitmix}.
Similar to ShapeShift, Alice sends $3$ BTC to deposit address \texttt{3Hp1Fk} generated by the service.
This input will be distributed to Bob with $2$ BTC as output and an temporary change address \#1 with $1$ BTC.
Then the balance of the address \#1 will be distributed to Charlie with $0.5$ BTC as output and another temporary change address \#2 with $0.5$ BTC.
The address \#2 is a special address that holds the remaining change after distributing user outputs and its balance is too small to enter the next round.
As a result, this trailing change will be consumed by the ending point of this chain.
This transaction consumes remaining changes from multiple peeling chains and merges them into a large balance for further use.

\subsection{Advanced Transaction Analysis}
\label{subsec:behavior_analysis}
As discussed in Section~\ref{subsec:behavioral_analysis_method}, mixing services using
the obfuscating mechanism allow us to identify more mixing transactions using a group of seeds. 
Therefore, Chipmixer and Wasabi Wallet can be further analyzed accordingly.

In the following, we first evaluate the
effectiveness of the proposed Algorithm~\ref{algo}. 
Due to the space limit, we only report the result for Chipmixer.
Then based on insights observed from identified transactions,
we are able to measure the profit made by each service.
Finally, we provide a case study to demonstrate the capability
of tracking the Bitcoin based on identified mixing transactions using
our proposed algorithm.
 
\subsubsection{Measuring the Effectiveness of the Algorithm}
Due to the lack of ground truth, we manually investigated our own ground truth to support the measurement. 
Specifically, for each service, we first collected transactions according to the common features we observed from the sample transactions and then filtered false positives manually. 
Then we were able to evaluate the robustness of the proposed algorithm by comparing the result with the ground truth.

\noindent
\textbf{Chipmixer.}
We conducted the following four experiments with different seeds (note that the $20$ sample transactions in Section~\ref{subsec:acquisition_result} are used as the original seed set).
\begin{itemize}
    \item \textit{Experiment 1.} We performed the first experiment at block height $609,750$.
    Using all mixing transactions identified in experiments as the
    seed set, we found $8,279$ transactions potentially generated by Chipmixer.
    \item \textit{Experiment 2.} We performed the second experiment at block height $619,700$,
    and found additional $1,056$ transactions ($9,335$ in total) mixing transactions from Chipmixer.
    \item \textit{Experiment 3.} We conducted the third experiment at the same block height with experiment 2. The seed set was randomly chosen from the original seed set with only half the size (10 transactions in total).
    We achieved the same expansion set as in experiment 2.
    \item \textit{Experiment 4.} We conducted the final experiment at the same block height. The seed set was only \textit{one} transaction randomly picked from the original seed set.
    Again, we achieved the same expansion set as in experiment 2.
\end{itemize}

These four experiments demonstrate that our method to identify mixing transactions is robust against \textit{different sizes} of the seed sets, and the same seed set $E$ can be used at \textit{different times} to identify mixing transactions from the same service. 
The summary of the experiments is shown in Table~\ref{table:cm_exp}.

\begin{table}[t]
\small
\caption{\label{table:cm_exp} Experiments to Evaluate the Seed-Expansion Algorithm for Chipmixer.}
\centering
\scalebox{0.9}{
\begin{tabular}{|c | c | c | c | c | c | c | c|  c|}
 \hline
  \textbf{Experiment} & \#1 & \#2 & \#3 & \#4 \\
  \hline
  \textbf{Date} & Dec 25, 2019 & Mar 1, 2020 & Mar 1, 2020 & Mar 1, 2020 \\
  \hline
  \textbf{Block Height} & 609,750 & 619,700 & 619,700 & 619,700 \\
  \hline
  \textbf{Seed Set} & 20 & 20 & 10 & \textbf{1} \\
  \hline
  \textbf{Expansion Set} & 8,279 & 9,335 & 9,335 & 9,335 \\
  \hline
  \textbf{Ground Truth} & 9,027 & 10,119 & 10,119 & 10,119 \\
  \hline
  \textbf{Coverage} & 91.71\% & 92.25\% & 92.25\% & 92.25\% \\
  \hline
  \textbf{Average Coverage} & \multicolumn{4}{c|}{\textbf{92.07\%}} \\
  \hline
\end{tabular}
}
\end{table}
 
\subsubsection{Calculating Profit of Mixing Services}
For each service, we will calculate the profit based on
identified mixing transactions.

\noindent
\textbf{Chipmixer.}
This service uses the Pay-What-You-Want (PWYW) pricing strategy (as described in Section~\ref{subsec:selection_result}), and will treat any change less than $0.001$ BTC as fees or donations \footnote{E.g., user input of $0.0015$ BTC will result in one chip with $0.001$ BTC (with $0.0005$ as the service fee), and an $0.0005$ BTC user input will be considered as fees or donations.}.

\begin{figure}[t]
	\centering
  \subfigure[Chipmixer.]{
      \includegraphics[width=.45\textwidth]{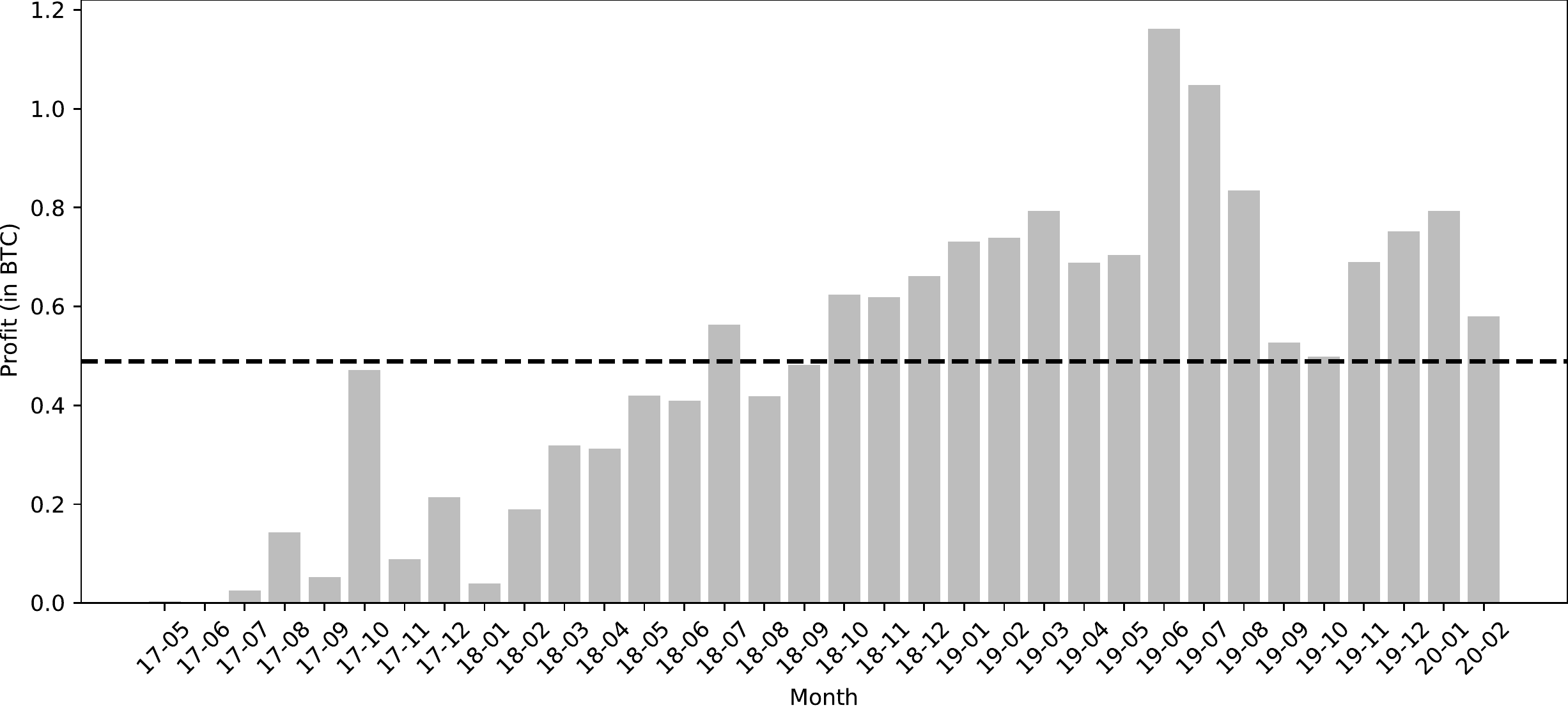}
      \label{fig:cm_profit}
  } \hfill
  \subfigure[Wasabi Wallet.]{
      \includegraphics[width=.45\textwidth]{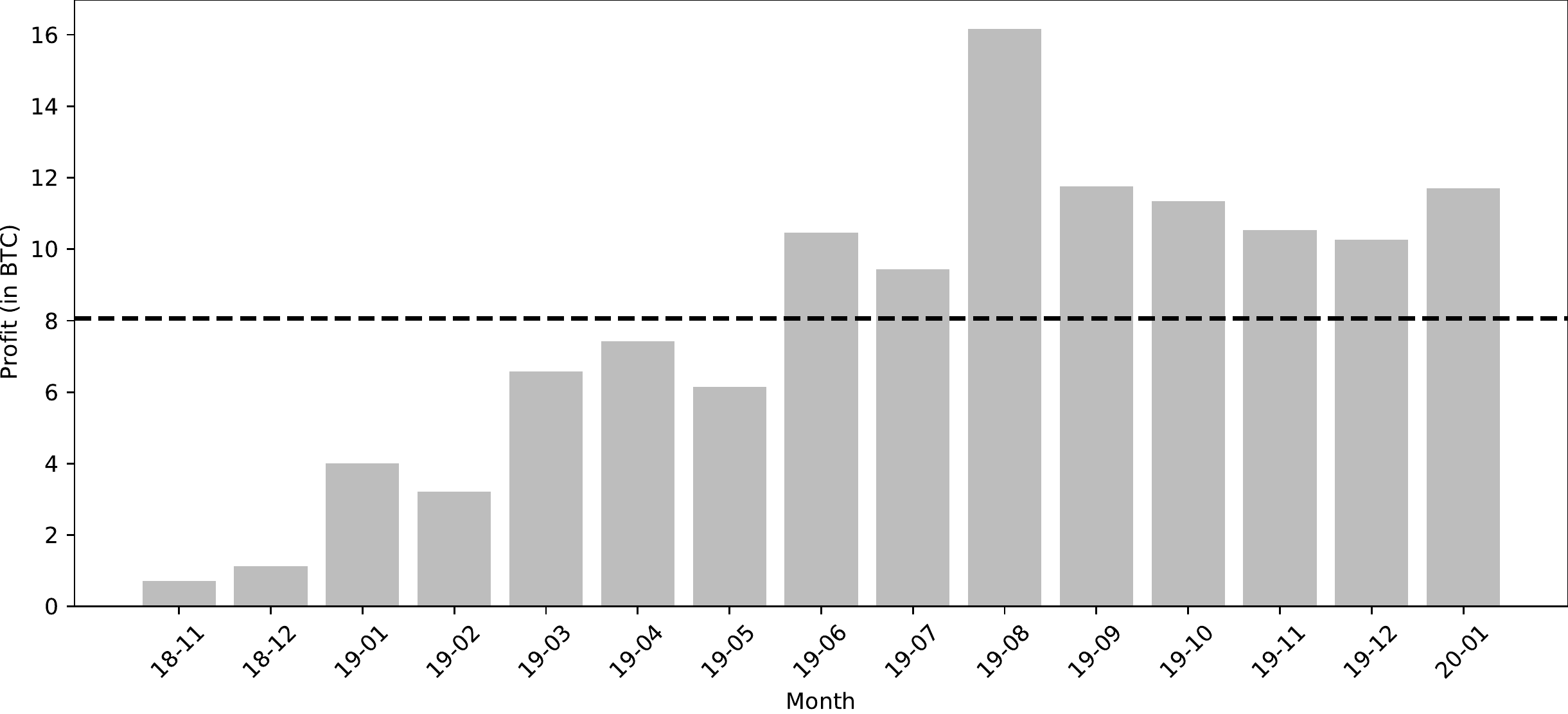}
      \label{fig:wasabi_profit}
  } 
\caption{Monthly profit for Chipmixer and Wasabi Wallet.
The dash line represents the average value.}
\vspace{-1em}
\label{fig:profits}

\end{figure}    

\label{subsubsec:attack}
\begin{figure*}[t]
	\centering
	\includegraphics[width=.9\textwidth]{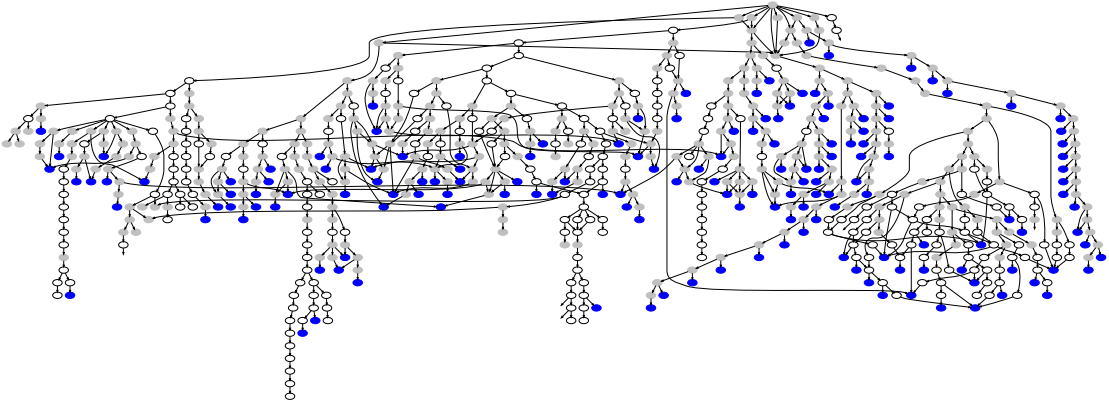}
	\caption{
		The simplified transaction graph for the Binance May Hack case.
		Transactions related to Chipmixer are annotated by blue nodes.
		In total, the attacker sent 4,792 BTC to Chipmixer.
	} 
	\label{fig:cm_case}
\end{figure*}

To calculate the profit earned by Chipmixer,
we sum over all trailing changes of user inputs from May, 2017 to February, 2020.
In total, Chipmixer received ${16.6086}$ BTC as service fees (with monthly average value ${0.4883}$ BTC), 
which was considerably less than the total user inputs $53,044.8077$ BTC during this period.
Figure~\ref{fig:cm_profit} illustrates monthly profit earned by Chipmixer.
Note that, our calculation for Chipmixer only serves as a lower bound.

\noindent
\textbf{Wasabi Wallet.}
We present the analysis based on the $9,788$ transactions obtained using the proposed Algorithm~\ref{algo}.
Similarly, our goal is to estimate the profit harvested by Wasabi Wallet for the CoinJoin fees.
As introduced in Section~\ref{subsubsec:results_interacting}, Wasabi Wallet's profit comes from the CoinJoin coordinate fees.
By analyzing every CoinJoin transaction identified, we found two common output addresses potentially for fee collection.
Address 1\footnote{Address: \texttt{bc1qs604c7jv6amk4cxqlnvuxv26hv3e48cds4m0ew}} has been used in $5,319$ CoinJoin transactions, but is no longer active since September 20, 2019.
Address 2\footnote{Address: \texttt{bc1qa24tsgchvuxsaccp8vrnkfd85hrcpafg20kmjw}} has been used in $3,204$ transactions and is currently active.
In every CoinJoin transaction, output value to these two addresses is close to the estimated coordinator fees.

Therefore, it is likely that these two addresses are used to collect coordinate fees.
Figure~\ref{fig:wasabi_profit} illustrates monthly profit earned by Wasabi Wallet.
In total, these addresses collected $120.9932$ BTC (with monthly average $8.058$ BTC), and it serves as a good estimation for the fees collected from Wasabi Wallet CoinJoin service.

\subsubsection{Tracing Money Flow of A Real Attack}
\label{subsubsec:case_study}
Finally, we demonstrate that our approach and results can help to reveal money laundering by tracing the money flow of stolen Bitcoins.

Specifically, we provide a simple case study for the Binance May Hack case~\cite{binancereport}. In this case, the attacker stole ${7,074}$ BTC and used Chipmixer for money laundering.
Starting from the attacker's output transaction \texttt{e8b406}, we track down the transaction graph to see whether any tainted funds are sent to Chipmixer. 
We use the identified transactions in previous experiments to test if a transaction sends Bitcoin to Chipmixer.
To solve the problem of dimension explosion, we set the maximum depth of tracing to 50 and ignore outputs less than $0.9$ BTC.

In total, we found $157$ transactions in identified transactions of Chipmixer, for a total value of ${4,797.82}$ BTC~\footnote{As a reference, an industry report~\cite{clain} gives an estimate of $4,836$ BTC were laundered through Chipmixer.}.
Figure~\ref{fig:cm_case} is a \textit{simplified} transaction graph to illustrate the case, where nodes are transactions. 
The blue nodes indicate transactions sending the tainted funds to Chipmixer,
while the gray ones mean that their addresses are \texttt{bc1q} addresses, which are coherent with the original outputs in transaction \texttt{e8b406}. 
Without the proposed approach, obviously, it may require a lot of human efforts to investigate the provenance of the stolen Bitcoins.
 
\section{Discussion}
\label{sec:discuss}
\noindent
\textbf{Threshold Parameter in Refined Algorithm.}\tab
In Section~\ref{subsec:acquisition_result}, we propose a refined version of the algorithm in~\cite{yousaf2019shapeshift}. 
It has a threshold parameter that limits the number of blocks to be examined. 
For the original algorithm, it is represented by two parameters ($\delta_a$ and $\delta_b$), which are determined by an optimization algorithm to examine 2 blocks in total ($\delta_a=1$, $\delta_b=0$, plus the block with the closest timestamp). 
However, in our evaluation it leads to poor performance ($80.29$\% records matched, compared with $90.29$\% of our refined algorithm).
After trying with different values, we manually set this parameter to examine 7 blocks in total, which is a trade-off between block coverage (larger threshold means more blocks examined) and performance (larger threshold means more false positives and less efficiency).
Obviously, our refinement leads to a much better performance.

\noindent
\textbf{Traceability Beyond Mixing Services.}\tab
Our approach only traces Bitcoins that are sent to mixing services. Tracing Bitcoins beyond mixing services is much more complicated because the money laundering may involve some off-chain activities (e.g., Over-The-Counter transactions) which cannot be traced through the on-chain information. However, our approach is still meaningful to serve many research and practical purposes (e.g., assisting criminal investigation involved with Bitcoins in Section~\ref{subsubsec:attack}). 

\noindent
\textbf{The Scope and Completeness of Our Study.}\tab
The scope of mixing services is limited due to the complexity of the ecosystem.
In this paper, we only consider traditional mixing services that fully rely on on-chain mechanisms to operate, without additional protocols.  
Other real-world mixing protocols like Fair Exchange and CoinSwap (investigated in~\cite{moser2017anonymous}), have much less popularity than traditional ones.
Besides, there exist complex research mixing protocols like Blindcoin~\cite{blindcoin} and Mixcoin~\cite{bonneau2014mixcoin}.
However, to the best of our knowledge, they have no real-world deployments.

\noindent
\textbf{Other Limitations of Our Work.}\tab
Our work has several limitations.
First, the advanced transaction analysis does not cover mixing services using
the swapping mechanism.
The design of peeling chains (see Section~\ref{subsubsec:swaapingmechanism}) deliberately hide mixing transactions by mimicking the features of normal user transactions.
We may have to seek other approaches to identify peeling chains and recover the relations of the transactions. 
Indeed, it is a technical challenge.

Another limitation arises from the two-step approach to identify anonymity sets in
the advanced transaction analysis (Section~\ref{subsec:behavioral_analysis_method}). If any output generated by a mixing transaction is incidentally not used as part of any transaction's inputs~\footnote{Or in some rare cases, these outputs are used as part of a transaction's inputs, but all the other parts do not belong to any other anonymity sets.}, then our approach could not find this mixing transaction.
Besides, it also relies on the size of the anonymity sets generated by mixing transactions. The smaller size will decrease the opportunity for outputs within the anonymity set to be used by other transactions as inputs, and thereby reducing the possibility of being identified.

In addition, as there does not exist any available data, we have to build the ground truth by ourselves. Although we have made our best efforts to eliminate the false positives, it inevitably may have some bias that affects the effectiveness of the measurement.
 
\section{Related Work}
\label{sec:relatedwork}
\noindent
\textbf{Bitcoin Mixing Service.}
The basic idea of mixing is to preserve relationship anonymity by obfuscating the relations from senders to recipients. 
Several mixing services have been publicly announced since 2010, including \textit{BitLaundry}~\cite{bitlaundry}, \textit{Bitcoin Laundry}~\cite{bitcoinlaundry} and \textit{Bitcoin Fog}~\cite{bitcoinfog}. In 2013, Maxwell made the idea of \textit{CoinJoin} public to the community~\cite{maxwell2013coinjoin}. In 2014, \textit{Mixcoin}~\cite{bonneau2014mixcoin} was proposed as the first academic work of mixing. Since then, a number of mixing approaches have been proposed, including \textit{Fair Exchange Protocol}~\cite{heilman2017tumblebit} and \textit{Zero Knowledge Proof}~\cite{zkcp}, and some of them have been implemented as services.
Generally speaking, there are mainly two types of mixing services, i.e., centralized (e.g., \textit{Bitcoin Fog}~\cite{bitcoinfog}, \textit{Mixcoin}~\cite{bonneau2014mixcoin} and \textit{Blindcoin}~\cite{blindcoin}) and decentralized (e.g., \textit{CoinJoin}~\cite{maxwell2013coinjoin}, \textit{CoinShuffle}~\cite{ruffingPedro2014esorics} and \textit{CloakCoin}~\cite{cloakcoin}).
The centralized mixing services rely on central mixing servers to perform mixing, while decentralized mixing services allow users to perform mixing without any centralized mixing server.
There are also centralized mixing services using decentralized protocols (like Wasabi Wallet using CoinJoin).
Besides, mixing services like ShapeShift~\cite{shapeshift} allow mixing across different blockchains.

\noindent
\textbf{Analyzing Bitcoin Mixing Service.}
Though mixing services have been widely used in the Bitcoin ecosystem, few studies have been published to understand them.
Möser et al.~\cite{moser2013mixing} conducted the first empirical study to analyze three Bitcoin mixing services focused on money laundering.
Yanovich et al.~\cite{bitfuryreport}
provided a heuristic-based algorithm to detect mixing transactions, and revealed that mixing transactions constituted about $2.5$\% of all transactions.
Balthasar et al.~\cite{de2017analysis} applied the tool provided by Chainalysis~\cite{chainalysis} to analyze three selected services and discovered severe security flaws in these services. 
However, their methods are specific to selected services and cannot be generalized to other mixing services.
Möser et al.~\cite{moser2017joinmarket} analysed the online \textit{CoinJoin} market named \textit{JoinMarket} and estimated its market volume.
Jaswant Pakki~\cite{new1pakki2020everything} provides a more recent survey on mixing services in Bitcoin, in which the author provides a table of mixing services with 9 trusted services.
Unlike these previous studies, we propose a generic model to systematically analyze state-of-the-art mixing services. 

\noindent
\textbf{Analyzing Raw Anonymity of Bitcoin.}
A number of research papers have been published to analyze raw anonymity properties of Bitcoin~\cite{khalilov2018survey} by either identifying the relations between Bitcoin addresses and user information,
or clustering and labeling Bitcoin addresses.
Our work is closed to those that mainly focused on Bitcoin addresses by analyzing blockchain data.
Reid et al.~\cite{reid2013analysis} proposed the first analytical results on the basis of two network structures, i.e., \textit{transaction network} and \textit{address network}, which can be used to depict money flow between transactions and users respectively. These two structures are widely used in subsequent researches~\cite{khalilov2018survey}.
Since then, several assumptions and methods were proposed and some of them have been used together to cluster Bitcoin addresses, including the multi-input heuristic~\cite{reid2013analysis,ron2013quantitative,androulaki2013evaluating,meiklejohn2013fistful,ober2013structure,spagnuolo2014fc,lischke2016analyze}, change addresses~\cite{androulaki2013evaluating,meiklejohn2013fistful,spagnuolo2014fc,neudecker2017fc} and behavior-based clustering~\cite{androulaki2013evaluating,ron2014fc}.
Although mixing services are rarely considered by these works, their methods and findings (e.g., the multi-input heuristic) form the basis of our work.
 
\section{Conclusion}
\label{sec:conclusion}
In this work, we aim to understand Bitcoin mixing services.
Accordingly, we first categorize mixing services into two types based on mixing mechanisms, i.e., swapping and obfuscating.
Then we propose a transaction analysis method to identify mixing mechanisms and workflows of these services.
Lastly, we propose a heuristic-based algorithm to identify mixing transactions.

We then apply the proposed approach to four representative mixing services.
The evaluation results demonstrate the effectiveness of our approach.
Specifically, we successfully determine the mixing mechanisms of each service.
We also show that it is able to identify most (over $92$\%) of the mixing transactions 
by applying the proposed algorithm. 
We finally provide two case studies, including calculating the profit and investigating the money laundering activity,
to show the usage scenarios of our study.
 
\section*{Acknowledgment}
The authors would like to thank the anonymous
reviewers for their insightful comments.
that helped improve the presentation of this paper.
This work was partially supported by the Fundamental Research Funds for the Central
Universities (No. 2020QNA5019, 2019QNA5016), Leading Innovative and Entrepreneur
Team Introduction Program of Zhejiang (No. 2018R01005),
the National Natural Science Foundation of China (grant No.62072046),
and Hong Kong RGC Projects (No. 152193/19E, 152223/20E).
Any opinions, findings,
and conclusions or recommendations expressed in this material
are those of the authors and do
not necessarily reflect the views of funding agencies.

\bibliographystyle{plain}
\bibliography{main}

\begin{thebibliography}{10}

\bibitem{bitcoinlaundry}
{Akemashite Omedetou}.
\newblock {Bitcoin Laundry}.
\newblock \url{https://en.bitcoin.it/wiki/Bitcoin\_Laundry}, 2011.

\bibitem{androulaki2013evaluating}
Elli Androulaki, Ghassan~O Karame, Marc Roeschlin, Tobias Scherer, and Srdjan
  Capkun.
\newblock Evaluating user privacy in bitcoin.
\newblock In {\em Proceedings of the International Conference on Financial
  Cryptography and Data Security}, 2013.

\bibitem{bartoletti2018data}
Massimo Bartoletti, Barbara Pes, and Sergio Serusi.
\newblock Data mining for detecting bitcoin ponzi schemes.
\newblock In {\em Proceedings of the Crypto Valley Conference on Blockchain
  Technology}, 2018.

\bibitem{binancereport}
Binance.
\newblock {Binance Security Breach Update}.
\newblock \url{https://www.binance.com/en/support/articles/360028031711}, 2019.

\bibitem{bistarelli2018visualizing}
Stefano Bistarelli, Matteo Parroccini, and Francesco Santini.
\newblock Visualizing bitcoin flows of ransomware: Wannacry one week later.
\newblock In {\em Proceedings of the Second Italian Conference on Cyber
  Security}, 2018.

\bibitem{bitcoinfog}
{Bitcoin Wiki}.
\newblock {Bitcoin Laundry}.
\newblock \url{https://bitcointalk.org/index.php?topic=50037}, 2011.

\bibitem{bitlaundry}
{Bitcoin Wiki}.
\newblock {BitLaundry}.
\newblock \url{https://en.bitcoin.it/wiki/BitLaundry}, 2011.

\bibitem{bitcointalk}
BitcoinTalk.
\newblock {Official Website of BitcoinTalk}.
\newblock \url{https://www.bitcointalk.org}, 2009.

\bibitem{bitmixbizann}
Bitmix.
\newblock {Announcement Thread of Bitmix.biz on BitcoinTalk}.
\newblock \url{https://bitcointalk.org/index.php?topic=2099519}, 2017.

\bibitem{bitmixbiz}
Bitmix.
\newblock {Official Website of Bitmix.biz}.
\newblock \url{https://bitmix.biz}, 2017.

\bibitem{blau2017price}
Benjamin~M. Blau.
\newblock Price dynamics and speculative trading in bitcoin.
\newblock {\em Research in International Business and Finance}, 2017.

\bibitem{bonneau2014mixcoin}
Joseph Bonneau, Arvind Narayanan, Andrew Miller, Jeremy Clark, Joshua~A Kroll,
  and Edward~W Felten.
\newblock Mixcoin: Anonymity for bitcoin with accountable mixes.
\newblock In {\em Proceedings of the International Conference on Financial
  Cryptography and Data Security}, 2014.

\bibitem{chainalysis}
{Chainalysis}.
\newblock {Official Portal of Chainalysis}.
\newblock \url{https://www.chainalysis.com/}, 2020.
\newblock (visited on 2020-05-21).

\bibitem{changelly}
Changelly.
\newblock {Official Website of Changelly}.
\newblock \url{http://changelly.com/}, 2015.

\bibitem{chipmixer}
Chipmixer.
\newblock {Official Website of Chipmixer}.
\newblock \url{https://chipmixer.com/}, 2017.

\bibitem{christin2013silkroad}
Nicolas Christin.
\newblock Traveling the silk road: A measurement analysis of a large anonymous
  online marketplace.
\newblock In {\em Proceedings of the 22nd International Conference on World
  Wide Web}, 2013.

\bibitem{clainreport}
Clain.
\newblock {Binance Hack} 2019.
\newblock
  \url{https://blog.clain.io/binance-hack-2019-\\deep-dive-into-the-money-laundering/},
  2019.

\bibitem{clain}
{Clain Team}.
\newblock {Binance Hack 2019 – A Deep Dive Into Money Laundering And Mixing}.
\newblock
  \url{https://blog.clain.io/binance-hack-2019-deep-dive-into-the-money-laundering/},
  2020.

\bibitem{cloakcoin}
{CloakCoin Official Portal}.
\newblock {CloakCoin}.
\newblock \url{https://www.cloakcoin.com/}, 2014.

\bibitem{coinmarketcap}
CoinMarketCap.
\newblock {Global Charts of CoinMarketCap}.
\newblock \url{https://coinmarketcap.com/charts/}, 2020.

\bibitem{de2017analysis}
Thibault de~Balthasar and Julio Hernandez-Castro.
\newblock An analysis of bitcoin laundry services.
\newblock In {\em Proceedings of the Nordic Conference on Secure IT Systems},
  2017.

\bibitem{bestmixertakedown}
Europol.
\newblock {Multi-Millon Euro Cryptocurrency Laundering Service BestMixer.io
  Taken Down}.
\newblock
  \url{https://www.europol.europa.eu/newsroom/news/multi-million-euro-cryptocurrency-laundering-service-bestmixerio-taken-down/},
  2019.

\bibitem{flypme}
Flyp.me.
\newblock {Official Website of Flyp.me}.
\newblock \url{https://flyp.me/en/}, 2012.

\bibitem{bitmixershutdown}
Rupert Hackett.
\newblock {BitMixer Shuts Down to ``Make Bitcoin Ecosystem More Clean''}.
\newblock
  \url{https://venturebeat.com/2017/07/25/bitmixer-shuts-down-to-make-bitcoin-ecosystem-more-clean/},
  2017.

\bibitem{harrigan2016clustering}
Martin Harrigan and Christoph Fretter.
\newblock The unreasonable effectiveness of address clustering.
\newblock In {\em Proceedings of the International IEEE Conferences on
  Ubiquitous Intelligence Computing, Advanced and Trusted Computing, Scalable
  Computing and Communications, Cloud and Big Data Computing, Internet of
  People, and Smart World Congress (UIC/ATC/ScalCom/CBDCom/IoP/SmartWorld)},
  2016.

\bibitem{heilman2017tumblebit}
Ethan Heilman, Leen Alshenibr, Foteini Baldimtsi, Alessandra Scafuro, and
  Sharon Goldberg.
\newblock Tumblebit: An untrusted bitcoin-compatible anonymous payment hub.
\newblock In {\em Proceedings of the 24th Annual Network and Distributed System
  Security Symposium}, 2017.

\bibitem{ibrahim2017ijns}
Maged~Hamada Ibrahim.
\newblock Securecoin: A robust secure and efficient protocol for anonymous
  bitcoin ecosystem.
\newblock {\em International Journal of Network Security}, 2017.

\bibitem{kethineni2018use}
Sesha Kethineni, Ying Cao, and Cassandra Dodge.
\newblock Use of bitcoin in darknet markets: Examining facilitative factors on
  bitcoin-related crimes.
\newblock {\em American Journal of Criminal Justice}, 2018.

\bibitem{khalilov2018survey}
Merve Can~Kus Khalilov and Albert Levi.
\newblock A survey on anonymity and privacy in bitcoinlike digital cash
  systems.
\newblock 2018.

\bibitem{helixclosedown}
{Larry Dean Harmon}.
\newblock {Helix Shutdown Announcement Thread on Reddit.com}.
\newblock \url{http://archive.fo/paKIO}, 2017.

\bibitem{lischke2016analyze}
Matthias Lischke and Benjamin Fabian.
\newblock Analyzing the bitcoin network: The first four years.
\newblock {\em Future Internet}, 2016.

\bibitem{maxwell2013coinjoin}
Gregory Maxwell.
\newblock {CoinJoin: Bitcoin} privacy for the real world.
\newblock \url{https://bitcointalk.org/index.php?topic=279249.0}, 2013.

\bibitem{zkcp}
{Maxwell, Gregory}.
\newblock {The first successful Zero-Knowledge Contingent Payment}.
\newblock
  \url{https://bitcoincore.org/en/2016/02/26/zero-knowledge-contingent-payments-announcement/},
  2016.

\bibitem{meiklejohn2013fistful}
Sarah Meiklejohn, Marjori Pomarole, Grant Jordan, Kirill Levchenko, Damon
  McCoy, Geoffrey~M Voelker, and Stefan Savage.
\newblock A fistful of bitcoins: characterizing payments among men with no
  names.
\newblock In {\em Proceedings of the Conference on Internet Measurement
  Conference}, 2013.

\bibitem{moser2017anonymous}
Malte M{\"o}ser and Rainer B{\"o}hme.
\newblock Anonymous alone? measuring bitcoin’s second-generation
  anonymization techniques.
\newblock In {\em Proceedings of the 2017 IEEE European Symposium on Security
  and Privacy Workshops}, 2017.

\bibitem{moser2017joinmarket}
Malte M{\"o}ser and Rainer B{\"o}hme.
\newblock The price of anonymity: empirical evidence from a market for bitcoin
  anonymization.
\newblock {\em Journal of Cybersecurity}, 2017.

\bibitem{moser2013mixing}
Malte M{\"o}ser, Rainer B{\"o}hme, and Dominic Breuker.
\newblock An inquiry into money laundering tools in the bitcoin ecosystem.
\newblock In {\em Proceedings of the 2013 APWG eCrime Researchers Summit},
  2013.

\bibitem{nakamoto2008bitcoin}
Satoshi Nakamoto.
\newblock Bitcoin: A peer-to-peer electronic cash system.
\newblock \url{https://bitcoin.org/bitcoin.pdf}, 2008.

\bibitem{neudecker2017fc}
Till Neudecker and Hannes Hartenstein.
\newblock Could network information facilitate address clustering in bitcoin.
\newblock In {\em Proceedings of the International Conference on Financial
  Cryptography and Data Security}, 2017.

\bibitem{noether2015monero}
Shen Noether.
\newblock Ring signature confidential transactions for monero.
\newblock https://eprint.iacr.org/2015/1098, 2015.

\bibitem{ober2013structure}
Micha Ober, Stefan Katzenbeisser, and Kay Hamacher.
\newblock Structure and anonymity of the bitcoin transaction graph.
\newblock {\em Future Internet}, 2013.

\bibitem{choosewallet}
Bitcoin Official.
\newblock {Choose Your Wallet}.
\newblock
  \url{https://bitcoin.org/en/choose-your-wallet?step=5&platform=windows},
  2017.

\bibitem{new1pakki2020everything}
Jaswant Pakki.
\newblock {\em Everything You Ever Wanted to Know About Bitcoin Mixers (But
  Were Afraid to Ask)}.
\newblock PhD thesis, Arizona State University, 2020.

\bibitem{pfitzmann2001term}
Andreas Pfitzmann and Marit K{\"o}hntopp.
\newblock Anonymity, unobservability, and pseudonymity—a proposal for
  terminology.
\newblock In {\em Proceedings of the International Workshop on Designing
  Privacy Enhancing Technologies: Design Issues in Anonymity and
  Unobservability}, 2001.

\bibitem{reid2013analysis}
Fergal Reid and Martin Harrigan.
\newblock An analysis of anonymity in the bitcoin system.
\newblock In {\em Proceedings of the 2011 IEEE Third International Conference
  on Privacy, Security, Risk and Trust and 2011 IEEE Third International
  Conference on Social Computing}. 2011.

\bibitem{ron2013quantitative}
Dorit Ron and Adi Shamir.
\newblock Quantitative analysis of the full bitcoin transaction graph.
\newblock In {\em Proceedings of the International Conference on Financial
  Cryptography and Data Security}, 2013.

\bibitem{ron2014fc}
Dorit Ron and Adi Shamir.
\newblock How did dread pirate roberts acquire and protect his bitcoin wealth?
\newblock In {\em Proceedings of the International Conference on Financial
  Cryptography and Data Security}, 2014.

\bibitem{ruffingPedro2014esorics}
Tim RuffingPedro and Moreno-SanchezAniket Kate.
\newblock Coinshuffle: Practical decentralized coin mixing for bitcoin.
\newblock In {\em Proceedings of the 19th European Symposium on Research in
  Computer Security}, 2014.

\bibitem{sasson2014zerocash}
Eli~Ben Sasson, Alessandro Chiesa, Christina Garman, Matthew Green, Ian Miers,
  Eran Tromer, and Madars Virza.
\newblock Zerocash: Decentralized anonymous payments from bitcoin.
\newblock In {\em Proceedings of the 2014 IEEE Symposium on Security and
  Privacy}, 2014.

\bibitem{shapeshift}
ShapeShift.
\newblock {Official Website of ShapeShift}.
\newblock \url{https://www.shapeshift.io}, 2014.

\bibitem{ssapirecenttx}
{ShapeShift}.
\newblock {\texttt{recenttx} API of Shapeshift}.
\newblock \url{http://shapeshift.io/recenttx/500}, 2020.

\bibitem{ssapitxstat}
{ShapeShift}.
\newblock {\texttt{txstat} API of Shapeshift}.
\newblock \url{https://shapeshift.io/txstat/[addr]}, 2020.

\bibitem{wsjshapeshift}
{Shifflett, Shane and Scheck, Justin}.
\newblock {The Wall Street Journal: How Dirty Money Disappears Into the Black
  Hole of Cryptocurrency}.
\newblock
  \url{https://www.wsj.com/articles/how-dirty-money-disappears-into-the-black-hole-of-cryptocurrency-1538149743},
  2019.

\bibitem{spagnuolo2014fc}
Michele Spagnuolo, Federico Maggi, and Stefano Zanero.
\newblock Bitiodine: Extracting intelligence from the bitcoin network.
\newblock In {\em Proceedings of the International Conference on Financial
  Cryptography and Data Security}, 2014.

\bibitem{bterloss}
{The Next Web}.
\newblock {Chinese Bitcoin exchange Bter will pay back users after losing
  \$1.75 million in cyberattack}.
\newblock
  \url{https://thenextweb.com/insider/2015/03/12/chinese-bitcoin-exchange-bter-will-pay-back-users-after-losing-1-75-million-in-cyberattack/},
  2015.

\bibitem{blindcoin}
Luke Valenta and Brendan Rowan.
\newblock Blindcoin: Blinded, accountable mixes for bitcoin.
\newblock In {\em Proceedings of the International Conference on Financial
  Cryptography and Data Security}, 2015.

\bibitem{wasabiapistates}
{Wasabi Wallet}.
\newblock {\texttt{states} API of Wasabi Wallet CoinJoin Service}.
\newblock \url{https://wasabiwallet.io/api/v3/btc/chaumiancoinjoin/states},
  2020.

\bibitem{wasabiapiunconfirmed}
{Wasabi Wallet}.
\newblock {\texttt{unconfirmed-coinjoins} API of Wasabi Wallet CoinJoin
  Service}.
\newblock
  \url{https://wasabiwallet.io/api/v3/btc/chaumiancoinjoin/unconfirmed-coinjoins},
  2020.

\bibitem{UTXO}
{Wikipedia}.
\newblock {Unspent transaction output}.
\newblock \url{https://en.wikipedia.org/wiki/Unspent_transaction_output}, 2020.

\bibitem{bitfuryreport}
Yuriy Yanovich, Pavel Mischenko, and Aleksei Ostrovskiy.
\newblock Shared send untangling in bitcoin.
\newblock
  \url{https://bitfury.com/content/downloads/bitfury_whitepaper_shared_send_untangling_in_bitcoin_8_24_2016.pdf},
  2016.

\bibitem{yousaf2019shapeshift}
Haaroon Yousaf, George Kappos, and Sarah Meiklejohn.
\newblock Tracing transactions across cryptocurrency ledgers.
\newblock In {\em Proceedings of the 28th USENIX Security Symposium}, 2019.

\bibitem{wasabiwallet}
zkSNACKs.
\newblock {Official Website of Wasabi Wallet}.
\newblock \url{https://wasabiwallet.io/}, 2017.

\end{thebibliography}

\end{document}